\documentclass[a4paper, 11pt]{article}

\usepackage{amsmath,amsfonts,amssymb,hyperref}

\usepackage{tikz}
\usepackage{color}
\usetikzlibrary{arrows}

\usepackage{graphicx} 
\usepackage{color}

\newcommand{\be}{\begin{equation}}
\newcommand{\ee}{\end{equation}}

\newcommand{\half}{\frac12}

\newcommand{\SL}{\mathrm{SL}}
\newcommand{\slrm}{\mathrm{sl}}

\newcommand{\Wcal}{\mathcal{W}}

\newcommand{\AdS}{\mathrm{AdS}}

\newcommand{\Cset}{{\,\,{{{^{_{\pmb{\mid}}}}\kern-.41em{\mathrm C}}}}}
\newcommand{\ket}[1]{\left|#1\right.\rangle}
\newcommand{\qb}{{\bar{q}}}

\newcommand{\p}{\partial}

\newcommand{\Lb}{\bar{L}}
\newcommand{\tb}{{\bar{t}}}

\newcommand{\comment}[1]{}

\begin{document}

\numberwithin{equation}{section}

\mbox{}

\vspace{40pt}

\begin{center}

{\Large \bf On the combinatorics of partition functions in $\mathrm{AdS}_3/\mathrm{LCFT}_2$}\\
\vspace{43pt}

{\large {\mbox{{\bf Yannick Mvondo-She$\,{}^a$} \hspace{.2cm} and \hspace{.2cm} {\bf Konstantinos Zoubos$\,{}^b$}}}}%
\vspace{.5cm}

Department of Physics, University of Pretoria\\
Private Bag X20, Hatfield 0028, South Africa

\mbox{}

and

\mbox{}

National Institute for Theoretical Physics (NITheP) \\
Gauteng, South Africa

\vspace{40pt}

{\Large \bf Abstract}

\end{center}

\noindent Three-dimensional Topologically Massive Gravity at its critical point has been conjectured to be holographically dual
  to a Logarithmic CFT. However, many details of this correspondence are still lacking. In this work, we study
  the 1-loop partition function of Critical Cosmological Topologically Massive Gravity,  previously derived
  by Gaberdiel, Grumiller and Vassilevich, and show that it can be usefully rewritten as a Bell polynomial expansion.
  We also show that there is a relationship between this Bell polynomial expansion and the plethystic exponential. Our
  reformulation allows us to  match the TMG partition function to states on the CFT side, including
  the multi-particle states of $t$ (the logarithmic partner of the CFT stress tensor) which had previously been elusive.
  We also discuss the appearance of a ladder action between the different multi-particle sectors in the partition function,
  which induces an interesting $\slrm(2)$ structure on the $n$-particle components of the partition function.

  \normalsize

\noindent

\vspace{4.1cm}
\noindent\rule{4.5cm}{0.4pt}

\noindent $^a$ vondosh7@gmail.com 

\noindent $^b$ kzoubos@up.ac.za

\vspace{0.5cm}

\setcounter{page}{0}
\thispagestyle{empty}
\newpage

\tableofcontents

\section{Introduction}

Three--dimensional gravity has, for quite some time now, served as an interesting setting in which
to test theories of gravity. Pure Einstein gravity in three dimensions is locally trivial at the
classical level, and does not exhibit propagating degrees of freedom. However, allowing for a
negative cosmological constant leads to a theory with black hole solutions \cite{Banados:1992wn} and
a careful study of the asymptotics \cite{Brown:1986nw} shows the emergence of a Virasoro algebra at the boundary.
One can thus expect a dual 2d CFT description, and this setting can be thought of as an early example of
the AdS/CFT correspondence \cite{Maldacena:1997re}.

One can also deform pure 3d gravity by adding a gravitational Chern-Simons term. This theory is
known as topologically massive gravity (TMG) and contains a massive graviton \cite{Deser:1982vy,Deser:1981wh}.
When both cosmological and Chern-Simons terms are included, the theory is known as cosmological topologically
massive gravity (CTMG). Such theories feature both gravitons and black holes.

Following Witten's proposal in 2007 to find a CFT dual to Einstein gravity \cite{Witten:2007kt}, the Einstein graviton 
1-loop partition function was calculated in \cite{Maloney:2007ud}. However, discrepancies were found 
in the results. In particular, the left- and right-moving contributions did not factorise, therefore clashing with the 
proposal of \cite{Witten:2007kt}.

Soon after, Li, Song and Strominger \cite{Li:2008dq} showed that the situation can be improved if one replaces
Einstein gravity  by \textit{chiral gravity}, which can be viewed as a special case of topologically massive gravity
\cite{Deser:1982vy, Deser:1981wh} at a specific tuning of the couplings ($\mu l=1$ in the notation of \cite{Li:2008dq},
where $l$ is the radius of $\AdS_3$
while $\mu$ is the coefficient of the gravitational Chern-Simons term), and is asymptotically defined
with Brown-Henneaux boundary conditions \cite{Brown:1986nw}. A particular feature of the theory was that one of
the two central charges vanishes: $c_L=0$, while $c_R\neq 0$. This gave an indication that the partition function could factorise.

Soon after the proposal of \cite{Li:2008dq}, based on the appearance of a non-trivial Jordan cell in CTMG 
at the critical point, the dual CFT at this point was conjectured to be logarithmic 
\cite{Grumiller:2008qz}. Indeed, Jordan cell structures are a salient feature of logarithmic CFTs (see 
\cite{Gurarie:1993xq}, as well as the very nice introductory notes \cite{Flohr:2001zs} and 
\cite{Gaberdiel:2001tr}). In the case of interest, the bulk Jordan cell structure arises between the
left-moving massless graviton mode and a massive bulk mode, which become degenerate at the critical point $\mu l=1$.
On the dual CFT side, this corresponds to the situation where the left-moving stress tensor $T$ acquires a logarithmic
partner state $t$, and we have the relations:
\be \label{Jordan}
L_0 \ket{T}=2\ket{T}\;\;,\;\; L_0 \ket{t}=2\ket{t}+\ket{T} \;\;,\;\;\Lb_0\ket{t}= T\;.
\ee
The proposal of \cite{Grumiller:2008qz} hinged on relaxing the Brown-Henneaux
boundary conditions in order to allow the presence of the logarithmic mode (see \cite{Grumiller:2008es, Henneaux:2009pw, Maloney:2009ck, Henneaux:2010fy} for discussions of the appropriate boundary conditions). This mode spoils the chirality
of the theory, as well as its unitarity. However, it opens up the very intriguing possibility of finding
bulk duals to logarithmic CFT's. A major milestone was the computation of correlation functions
\cite{Skenderis:2009nt,Grumiller:2009mw} in TMG, which confirmed the existence of logarithmic correlators
of the general type $\langle T(x) t(y)\rangle=b_L/(x-y)^4$, where $b_L$ is often called the logarithmic anomaly. For TMG it
takes the value $b_L=-3l/G_N$. 

In order to further understand the TMG/LCFT proposal\footnote{In the rest of this work
  we will refer to critical CTMG as simply TMG.}, the calculation of the 1-loop graviton partition function of TMG on
the thermal $\AdS_3$ background was 
undertaken in \cite{Gaberdiel:2010xv}, by means of heat kernel techniques such as in \cite{Giombi:2008vd, Gopakumar:2011qs}.
The partition function was found to be:\footnote{This is not a fully modular invariant partition function, but
 corresponds to $Z_{0,1}$ in the notation of \cite{Maloney:2007ud}.}
\be  \label{ZTMG}
{Z_{TMG}} (q, \bar{q})= \prod_{n=2}^{\infty} \frac{1}{|1-q^n|^2} \prod_{m=2}^{\infty} 
\prod_{\bar{m}=0}^{\infty} \frac{1}{1-q^m \bar{q}^{\bar{m}}}. 
\ee
 As shown in \cite{Gaberdiel:2010xv}, the partition function (\ref{ZTMG}) is consistent with an explicit
counting of states in the dual LCFT, up to the level of single-particle states in the logarithmic partner $t$.
This certainly strengthens the case for an LCFT dual of TMG. However,
as was also pointed out in \cite{Grumiller:2013at}, one would still desire a better understanding of the
partition function from the CFT side. In particular, although the multi-particle states encoded
in (\ref{ZTMG}) were consistent with those of a dual CFT (in that they have positive multiplicities), it
would be desirable to match them exactly with the combinatorics of the multi-particle excitations of the
logarithmic partner. This is the main question we consider in this work. 

The first step in understanding this counting will be to show how the partition function (\ref{ZTMG})
can be recast in terms of \emph{Bell polynomials}. The latter are very useful in many areas of mathematics
and have enjoyed many applications in physics as well. In our case, the Bell polynomial reformulation allows us to
explicitly match each term in the partition function to the descendants of multi-particle states of the logarithmic
partner. We also show that the
Bell polynomial form of the partition function is perfectly consistent with plethystic exponential
techniques \cite{Benvenuti:2006qr,Feng:2007ur}.  Finally, we discuss a ladder construction which
generates the $n$-particle part of the partition function from the $n-1$-particle part (and the reverse),
which also allows us to uncover an $\slrm(2)$ symmetry of the partition function. 

There also exist other topologically massive gravity theories, such as the non-chiral New Massive Gravity (NMG) \cite{Bergshoeff:2009hq},
which have been argued to have LCFT duals \cite{Grumiller:2009sn,Alishahiha:2010bw}. The partition 
function of NMG was also given in \cite{Gaberdiel:2010xv}. Furthermore, one can extend TMG by including
higher spin fields, which appears to lead to a dual LCFT with $\Wcal$-algebra symmetry \cite{Bagchi:2011vr}. 1-loop
partition functions for these higher-spin theories have been computed in \cite{Bagchi:2011td}. We show that
the partition functions of NMG and Higher-spin TMG can also be straightforwardly brought into Bell polynomial form.

This paper is organised as follows: in section 2, the combinatorial properties of Z$_{{TMG}}$ are 
derived in terms of Bell polynomials.  In section 3, the connection between the Bell polynomials and the plethystic 
exponential is developed. Section 4 illustrates how the matching of the terms in the partition
function to explicitly constructed multi-particle states in the CFT works. In section 5, we construct
ladder-type (raising and lowering) operators acting on the Bell polynomials and the 
plethystic exponential. These operators reveal an $\slrm(2)$ symmetry acting between the $n$-particle components
of the partition function. Finally, after briefly discussing the extension of our results to NMG and higher spin TMG
in section 6, we conclude with some open questions.

\section{Combinatorial properties of Z$_{{TMG}}$}

In this section, after briefly reviewing the results of \cite{Gaberdiel:2010xv} as well as some
relevant aspects of Bell polynomials, we show how $Z_{TMG}$ can be written as a Bell polynomial expansion.

\subsection{Matching of the vacuum and single-$t$ states} \label{singlet}

As shown in \cite{Gaberdiel:2010xv}, the partition function (\ref{ZTMG})  has 
the general structure expected from a dual logarithmic 
conformal field theory.
In order to see this, one can expand (\ref{ZTMG}) in two parts:
\begin{equation} \label{ZTMG0mp}
  {{Z_{TMG}}}(q, \bar{q})    = {{Z^0_{LCFT}}} (q, \bar{q}) + \sum_{h,\bar{h}} N_{{h,\bar{h}}} q^h 
\bar{q}^{\bar{h}} \prod_{n=1}^{\infty} \frac{1}{|1-q^n|^2}, 
\end{equation} 
with   
\begin{equation} \label{ZLCFT0}
{{Z^0_{LCFT}}(q, \bar{q})= Z_{\Omega} + Z_t} = \prod_{n=2}^{\infty} \frac{1}{|1-q^n|^2} \left( 1+ 
\frac{q^2}{|1-q|^2}\right) \;,
\end{equation} 
where $\Omega$ indicates the vacuum of the holomorphic sector, and $t$ denotes the logarithmic partner of the 
energy momentum tensor $T$.\footnote{Note that since the vacuum $\ket{\Omega}$ is $\SL(2)$ invariant, its
  descendants start at level 2, i.e. from the
state $L_{-2}\ket{\Omega}$. On the other hand, $Z_t$ also includes states such as $L_{-1}\ket{t}$. The holomorphic
dimension of $t$ is 2 (\ref{Jordan}), hence the $q^2$.} 
The coefficients $N_{h,\bar{h}}$ are higher-order in $h,\bar{h}$ and
should correspond to multi-particle states in $t$. Crucially, although a full understanding of the combinatorics
leading to the multi-particle part of (\ref{ZTMG0mp}) was not attempted  
in \cite{Gaberdiel:2010xv}, the coefficients $N_{h,\bar{h}}$ were shown to be all positive, as should be the
case if they are indeed counting states in a CFT. 

The above result was generalised to NMG as \cite{Gaberdiel:2010xv}:    
\be \label{ZNMG} 
{{Z_{NMG}}}(q, \bar{q})= \prod_{n=2}^{\infty} \frac{1}{|1-q^n|^2} \prod_{m=2}^{\infty} 
\prod_{\bar{m} =0}^{\infty} \frac{1}{1-q^m \bar{q}^{\bar{m}}} \prod_{l=0}^{\infty} 
\prod_{\bar{l}=2}^{\infty} \frac{1}{1-q^l \bar{q}^{\bar{l}}}. 
\ee
This can also be split into two parts, 
\be
  {{Z^{NMG}_{LCFT}}}(q, \bar{q})= {{Z^{(0)NMG}_{LCFT}}}(q, \bar{q}) + \sum_{h,\bar{h}} 
N_{{h,\bar{h}}} q^h \bar{q}^{\bar{h}} \prod_{n=1}^{\infty} \frac{1}{|1-q^n|^2}, 
\ee
with
\be \label{Z0NMG}
{{Z^{(0)NMG}_{LCFT}}}(q, \bar{q}) = Z_{\Omega} + Z_{t}+Z_{\bar{t}}  = \prod_{n=2}^{\infty} 
\frac{1}{|1-q^n|^2} \left( 1 + \frac{q^2 + \bar{q}^2}{|1-q|^2}\right). 
\ee
We see that, as before, the partition function can be written as the descendants of the vacuum, as well
as terms corresponding to the descendants of the left- and right- logarithmic partners $t$ and $\bar{t}$.

We recall that the CFT dual to TMG is expected to have logarithmic behaviour in the holomorphic sector, with $c_L=0$,
with the antiholomorphic sector being non-logarithmic and with $c_R\neq 0$. On the other hand,
for the  LCFT dual to NMG both sectors are expected to be logarithmic, with $c_L=c_R=0$. The bulk mode dual
to $t$ results in the
single double product appearing in (\ref{ZTMG}), while the bulk modes dual to $t$ and $\bar{t}$ (the logarithmic
partner of the right-moving CFT stress tensor) lead to the
two double products appearing in (\ref{ZNMG}). In what 
follows, we will be interested in better understanding the structure of such double products,
focusing mainly on the TMG case.   

\subsection{Multipartite generating functions} 

Our main tool for the study of the multi-particle terms in $Z_{TMG}$ will be multipartite generating
functions. Following e.g. \cite{theoryofpartitions} (see also \cite{Bytsenko:2017czh} for a recent summary of the technique and
applications), let us review how multipartite generating functions can be written in terms of Bell polynomials,
a result also known as the Fa\`{a} di Bruno formula.   

For any \emph{multipartite} (or $m$-partite) number $\vec{k}=(k_1,k_2, \ldots, k_m)$, i.e. any ordered
$m$-tuple of non-negative integers not all zeros, let $N^{(z;m)}(\vec{k})= N^m(z;k_1,k_2, \ldots,k_m)$
be the number of partitions of $\vec{k}$, i.e. the number of distinct representations of $(k_1,k_2, \ldots, k_m)$ as a sum
of multipartite numbers. The generating function of $N^{(z;m)}(\vec{k})$ is defined as: 
\begin{eqnarray} \label{generating}
G(z;X) = \prod_{\vec{k}\geq 0} \frac{1}{1-z x_1^{k_1} x_2^{k_2} \cdots x_m^{k_m}} = 
\sum_{\vec{k}\geq 0} N^{(z;m)}(\vec{k}) ~x_1^{k_1} x_2^{k_2} \cdots x_m^{k_m}\;. 
\end{eqnarray}

It follows that:   
 
\begin{eqnarray} 
\log G(z;X) &=& - \sum_{\vec{k}\geq 0} \log (1-z x_1^{k_1} x_2^{k_2} \cdots x_m^{k_m}) \\ \nonumber 
&=& \sum_{\vec{k}\geq 0} \sum_{n=1}^{\infty} \frac{z^n}{n} x_1^{nk_1} x_2^{nk_2} \cdots x_m^{nk_m} 
\nonumber \\ &=& \sum_{n=1}^{\infty} \frac{z^n}{n} \frac{1}{1-x^n_1} \frac{1}{1-x^n_2} \cdots \frac{1}{1-x^n_m} \nonumber \\ &=& \sum_{n=1}^{\infty} \frac{z^n}{n} \prod_{j=1}^m \frac{1}{1-x^n_j}, 
\end{eqnarray} 
and finally: 
\begin{eqnarray} 
\sum_{\vec{k}\geq 0} N^{(z;m)}(\vec{k}) x_1^{k_1} x_2^{k_2} \cdots x_m^{k_m} = \exp \left( 
\sum_{n=1}^{\infty} \frac{z^n}{n} \mathcal{F}_m(n) \right), 
\end{eqnarray} 
with $\mathcal{F}_m(n) = \prod \limits_{j=1}^m \frac{1}{1-x^n_j}$.

\subsection{The Fa\`{a} di Bruno formula and Bell polynomials} 

Bell polynomials were defined by E.T. Bell in 1934 \cite{bell1934exponential}, but their name is due to Riordan
\cite{riordan58} who studied the Fa\`{a} di Bruno formula \cite{faa1855sullo,di1857note} expressing 
the $n$-th derivative of a composite function $f \circ g$ in terms of the derivatives of $f$ and $g$ 
\cite{riordan1946}. Defining the shorthand notation: $d^n h / dx^n = h_n$, $d^n f / dg^n = f_n$ and $d^ng / dt^n = g_n$,
it is easy to see that
\be
h_1=f_1\;, \quad h_2= f_1 g_2 + f_2 g_1^2\,\quad h_3 = f_1 g_3 + 3 f_2 g_2 g_1 + f_3 g_1^3 \;,\dots
\ee
Using mathematical induction, one finds
\be
h_n = f_1 \beta_{n1}(g_1, \ldots, g_n) + f_2 \beta_{n2}(g_1, \ldots, g_n) + \ldots + f_n \beta_{nn}(g_1, \ldots, g_n) \;
\ee
where $\beta_{nj}(g_1, \ldots, g_n) $ is a homogeneous polynomial of degree $j$ in $g_1, \ldots, g_n$. 

It turns out that the study of $h_n$ simply reduces to the study of the \textit{Bell polynomials}: 
\be
Y_n(g_1, g_2, \ldots, g_n)= \beta_{n1}(g_1, \ldots, g_n) +    \beta_{n2}(g_1, \ldots, g_n) + \ldots +   
\beta_{nn}(g_1, \ldots, g_n)\;.
\ee
Beyond multipartite partition problems \cite{theoryofpartitions} these well-known polynomials
find applications in other aspects of combinatorics, number theory, analysis, probability and algebra.

A useful recurrence relation for the Bell polynomials $Y_n(g_1, g_2, \ldots, g_n)$  is  \cite{theoryofpartitions}:  
\be
Y_{n+1}(g_1, g_2, \ldots, g_{n+1}) = \sum^n_{k=0} \begin{pmatrix} n \\ k \end{pmatrix} Y_{n-k}(g_1, g_2, 
\ldots, g_{n-k}) g_{k+1}, 
\ee
We can also write a generating function  $\mathcal{G}(z)$:
\be 
\mathcal{G}(z)= \sum^{\infty}_{n=0} \frac{Y_n z^n}{n!} \hspace{0.5cm} \Rightarrow \hspace{0.5cm} 
\log \mathcal{G}(z)= \sum^{\infty}_{n=0} \frac{g_n z^n}{n!}. 
\ee 
From the above, one obtains the following explicit expression for the Bell polynomials (also referred to as 
the Fa\`{a} di Bruno formula):
\begin{eqnarray} \label{FaadiBruno}
Y_{n}(g_1, g_2, \ldots, g_{n})= \sum_{\vec{k} \vdash n} \frac{n!}{k_1! \cdots k_n!} \prod_{j=1}^n \left( 
\frac{g_j}{j!} \right)^{k_j}. 
\end{eqnarray} 
Here the notation {$\vec{k} \vdash n$} is defined as: 
\be \label{kdashn}
{\vec{k} \vdash n }= \{ {\vec{k}= ( k_1,k_2, \ldots,k_n ) \hspace{0.25cm} | \hspace{0.25cm} k_1 + 2k_2 + 
3k_3 + \cdots + nk_n=n} \}. 
\ee
In Appendix \ref{AppBell} we show how this formula can be expanded to give the explicit Bell polynomials, and also
briefly explain their combinatoric meaning.

\subsection{$Z_{TMG}$ in terms of Bell polynomials}

After these preliminaries, we are ready to show that ${Z_{TMG}}$ can be rewritten as an 
(exponential) generating function of Bell polynomials. We start by rewriting eq. (\ref{ZTMG}) as:   
\be\label{ZTMG1}
{Z_{TMG}}(q, \bar{q})= \prod_{n=2}^{\infty} \frac{1}{|1-q^n|^2} 
\prod_{m={\color{red}{0}}}^{\infty} \prod_{\bar{m}=0}^{\infty} \frac{1}{1- {\color{red}{q^2}} q^m \bar{q}^{\bar{m}}} \;,
\ee 
where we have just specialised $z$ in (\ref{generating}) to be $q^2$. Then, if we 
write: 
\begin{eqnarray} \label{ZTMGsplit}
{{Z_{TMG}}} = \mathcal{A}(q,\bar{q}) \mathcal{B}(q,\bar{q}), 
\end{eqnarray} 
with
\begin{eqnarray} 
\mathcal{A}(q,\bar{q}) = \prod_{n=2}^{\infty} \frac{1}{|1-q^n|^2}    \quad \text{and}\quad
\mathcal{B}(q,\bar{q}) = \prod_{m=0}^{\infty} \prod_{\bar{m}=0}^{\infty} \frac{1}{1- q^2 q^m 
\bar{q}^{\bar{m}}}, 
\end{eqnarray}
and focus on $\mathcal{B}(q,\bar{q})$, we compute: 
\begin{eqnarray} 
\log \mathcal{B}(q,\bar{q}) &=& - \sum_{m \geq 0, \bar{m} \geq 0} \log (1-q^2 q^m q^{\bar{m}}) 
\nonumber \\ &=& - \sum_{m \geq 0, \bar{m} \geq 0} \left( - \sum_{n=1}^{\infty} 
\frac{(q^2)^n}{n} q^{nm} \bar{q}^{n \bar{m}} \right) \nonumber \\ &=&   
\sum_{m \geq 0, \bar{m} \geq 0} \sum_{n=1}^{\infty} \frac{q^{2n}}{n} q^{nm} \bar{q}^{n \bar{m}} 
\nonumber \\ &=& \sum_{n=1}^{\infty}    \frac{q^{2n}}{n} \left( \sum_{m \geq 0, \bar{m} \geq 0}   
q^{nm} \bar{q}^{n \bar{m}} \right) \nonumber \\ &=& \sum_{n=1}^{\infty}    \frac{q^{2n}}{n!} \left[ 
(n-1)! \sum_{m \geq 0, \bar{m} \geq 0}    q^{nm} \bar{q}^{n \bar{m}} \right] \nonumber \\    &=& 
\sum_{n=1}^{\infty}    \frac{q^{2n}}{n!} g_n. 
\end{eqnarray}
Hence, we can write:  
\begin{eqnarray} \label{Bpart}
\mathcal{B}(q,\bar{q}) = \sum_{n=0}^{\infty} \frac{Y_n}{n!} q^{2n},   
\end{eqnarray} 
with $Y_{n}(g_1, g_2, \ldots, g_{n})$ defined as in (\ref{FaadiBruno})    
and 
 \be \label{gn} 
g_n = (n-1)! \sum_{m \geq 0, \bar{m} \geq 0}    q^{nm} \bar{q}^{n \bar{m}}. 
\ee 
For easy reference, the first few terms in the expansion (\ref{Bpart}) are given in appendix \ref{AppBell}. One finds
\be
\begin{split}
\mathcal{B}(q,\bar{q}) &= \frac{1}{0!} Y_0 \left(    q^2 \right)^0 +   
\frac{1}{1!} Y_1 \left(    q^2 \right)^1 + \frac{1}{2!} Y_2 \left(    q^2 \right)^2 + \frac{1}{3!} Y_3 \left(   
q^2 \right)^3 + \ldots  \\ 
&= 1+ Y_1 \left(    q^2 \right) + \frac{1}{2!} Y_2 \left(    q^2 \right)^2 + \frac{1}{3!} Y_3 \left(    q^2 
\right)^3 + \ldots, 
\end{split}
\ee 
where
\be
\begin{split}
Y_1 &= g_1 = \sum_{m \geq 0} \sum_{\bar{m} \geq 0}    q^{m} \bar{q}^{ \bar{m}}\;,\\ 
Y_2 &= g_1^2 + g_2 = \left( \sum_{m \geq 0} \sum_{\bar{m} \geq 0}    q^{m} \bar{q}^{ \bar{m}}   
\right)^2 +    \sum_{m \geq 0} \sum_{\bar{m} \geq 0}    q^{2m} \bar{q}^{2 \bar{m}} \\   
&=    \sum_{m \geq 0} \sum_{\bar{m} \geq 0} (m+1)(\bar{m}+1) q^{m} \bar{q}^{ \bar{m}} +    \sum_{m 
\geq 0} \sum_{\bar{m} \geq 0}    q^{2m} \bar{q}^{2 \bar{m}}\;,\\ 
Y_3 &= g_1^3 + 3 g_1 g_2 + g_3 \\ &= \left( \sum_{m \geq 0} \sum_{\bar{m} \geq 0}    q^{m} 
\bar{q}^{ \bar{m}}    \right)^3     
+ 3 \left( \sum_{m \geq 0} \sum_{\bar{m} \geq 0}    q^{m} \bar{q}^{ \bar{m}}    \right)    \left( \sum_{m 
\geq 0} \sum_{\bar{m} \geq 0}    q^{2m} \bar{q}^{2 \bar{m}}    \right) \\ &+ 2 \sum_{m \geq 0} 
\sum_{\bar{m} \geq 0}    q^{3m} \bar{q}^{3 \bar{m}}\;,
\end{split}
\ee
and so on. Recalling the form (\ref{ZTMGsplit}) of the partition function, we can now write
\be \label{ZTMGBellfinal}
Z_{TMG}(q,\qb)=\prod_{n=2}^\infty \frac{1}{|1-q^n|^2} \sum_{k=0}^\infty \frac{Y_{k}}{k!}q^{2k}\;.
\ee
We conclude that the double-product part of $Z_{TMG}$ in (\ref{ZTMG}) is an exponential generating series
of Bell polynomials. Expressing these polynomials in terms of the $g_n$ functions (\ref{gn}), we can
write out the first few terms in the expansion:
 \be \label{ZTMGexpanded}
{{Z_{TMG}}}(q, \bar{q}) = \prod_{n=2}^{\infty} \frac{1}{|1-q^n|^2} \left[    1+ \frac{g_1}{1!} q^2      + 
\frac{g_1^2 + g_2}{2!} \left(    q^2 \right)^2 + \frac{g_1^3 + 3 g_1 g_2 + g_3}{3!} \left(    q^2 \right)^3 
+\cdots \right].
\ee 
As we will explain in the next section, the usefulness of this reformulation is that each term within the square brackets corresponds
to a different $n$-particle sector of the logarithmic partner state $t$. The term arising from 
$Y_1$ contains the descendants of $t$, the term arising from $Y_2$ the descendants of $t\otimes t$ and so on.
So re-expressing the partition function as (\ref{ZTMGexpanded}) will make it easy to compare it to an explicit
construction of the descendants.

\subsection{The Bell series as a multi-particle generating function} \label{Bellmultiparticle}

Let us now provide an interpretation of the above combinatorial results from the LCFT perspective. In 
particular, we would like to suggest that in the expansion of $\mathcal{B}(q, \bar{q})$, while the terms 
$(q^2)^n$ correspond to single-particle and multi-particle states ($t$ and $t \otimes_n t$ 
respectively), the terms $Y_n$ for $n \geq 1$ are character representations of descendants of 
 $t$ when $n=1$, and of $t \otimes_n t$ for $n \geq 2$.

We start by recalling the well-known identity for geometric sums: 
\be
\sum_{n=0}^{\infty} q^n = \frac{1}{1-q}. 
\ee 
From the above equation, it is easy to see that: 
\be
\begin{split}
  g_n &= (n-1)! \sum_{m \geq 0, \bar{m} \geq 0}    q^{nm} \bar{q}^{n \bar{m}} 
  = (n-1)! \left(\sum^{\infty}_{m=0} \left( q^n \right)^m \right) \left( \sum^{\infty}_{\bar{m}=0} \left( \bar{q}^n 
  \right)^{\bar{m}} \right) \\
  &= (n-1)! \left( \frac{1}{1-q^n} \right) \left( \frac{1}{1-\bar{q}^n} \right)  
= (n-1)! \frac{1}{|1-q^n|^2} \;.
\end{split}
\ee
This allows us to rewrite (\ref{ZTMGexpanded}) up to third order in the expansion of $\left( q^2 \right)$ as: 
\be  \label{ZTMGexpandedfinal}
\begin{split}
{{Z_{LCFT}}}(q, \bar{q}) &=\prod_{n=2}^{\infty} \frac{1}{|1-q^n|^2}   
\left\{ 1+ \frac{\left( q^2 \right)^1}{|1-q|^2} + \frac{1}{2!} \left[ \left( \frac{1}{|1-q|^2} \right)^2 + 
  \frac{1}{|1-q^2|^2} \right] \left( q^2 \right)^2 \right. \\
&+ \left. \frac{1}{3!} 
\left[ \left( \frac{1}{|1-q|^2} \right)^3 + 3 \frac{1}{|1-q|^2}    \frac{1}{|1-q^2|^2} +
  \frac{2}{|1-q^3|^2}    \right] \left( q^2 \right)^3 + \cdots         
\right\}\;,
\end{split}
\ee 
where we now write $Z_{LCFT}$ to emphasise that we would like to interpret it as an LCFT partition function.
One can immediately see that the first two terms in the above expansion are the same as those in (\ref{ZLCFT0}).
As explained in \cite{Gaberdiel:2010xv}, the ``1'' in the braces times the overall prefactor corresponds to the descendants of the vacuum,
where we note that due to $\SL(2)$ invariance we have $L_{-1}\ket{\Omega}=\Lb_{-1}\ket{\Omega}=0$, so the descendants encoded in the
prefactor are all the states created by products of the $L_{-k}$ and $\Lb_{-\bar{k}}$, with $k,\bar{k}\geq 2$.

As also shown in \cite{Gaberdiel:2010xv}, the next term, proportional to $q^2$, corresponds to the descendants of the
logarithmic partner $t$. Let us note that the term inside the braces contains only the global $\SL(2)\times\SL(2)$ descendants of type $L_{-1}^k \ket{t}$ and
$\Lb_{-1}^{\bar{k}} \ket{t}$, while the prefactor provides the remaining Virasoro descendants. Although for the single-particle $t$ states
this is just a trivial rewriting, this observation will be crucial in understanding the structure of the multi-particle states. 

Before proceeding to the study of the multi-particle states of $t$, let us comment on our expectations regarding such states
in the context of the proposed TMG/LCFT correspondence. In a generic CFT, such states correspond to operators
appearing in the (iterated) OPE of $t$ with itself. Identifying these states would then require knowledge of the fusion rules
of the theory, which, lacking a concrete CFT construction, we do not have in this case. However, here we are assuming that the
LCFT has a gravity dual, which,
similarly to the study of \cite{Gaberdiel:2011zw} for the higher-spin minimal models, implies that the fusion rules should reduce to
just tensor product rules (note that, although in our case $c_L=0$, the logarithmic anomaly $b_L=-3l/G_N$ is large and provides the
``large $N$'' parameter).
  In particular, as in \cite{Gaberdiel:2011zw}, we will assume that the conformal weights of these tensor product states are
  additive, as a minimal requirement for a match to the multiparticle states on the gravity side to be possible.\footnote{A potential subtlety,
  observed in \cite{Gaberdiel:2011zw}, is that the state with additive quantum numbers might actually be a descendant of the tensor
  product state, with the tensor product state itself becoming null in the large-$N$ limit. However, this should have no effect on the counting,
  which would then start from the additive state.}
  This reasoning implies that a $k$-particle state of $\ket{t}$ with $(L_0,\Lb_0)$ eigenvalues $(2k,0)$ exists with multiplicity 1,
  and we will label it
  $\ket{t}\otimes_k \ket{t}$. This explains the $q^{2k}$ term in (\ref{ZTMGBellfinal})\footnote{Note that $Y_k/k!=1+\cdots$ so in the expansion (\ref{ZTMGBellfinal}) there is a $q^{2k}$ term with coefficient $1$ for all $k$.} and our task in the following
  will be to understand the
  combinatorics leading to the additional contributions coming from $Y_k/k!$  and the overall prefactor, which we will identify with the
  descendants of $\ket{t}\otimes_k \ket{t}$. 

   To this end, let us now proceed to consider the third term in the braces in (\ref{ZTMGexpandedfinal}), which comes
  from the $Y_2$ term in the Bell expansion. We claim that this term corresponds
to the descendants of the two-particle state $\ket{t}\otimes \ket{t}$, with the term in the  braces counting the contributions of
only the global $\SL(2)\times \SL(2)$ ($L_{-1}$ and $\Lb_{-1}$) descendants, while the overall prefactor counts the
$L_{-k}$ and $\Lb_{-\bar{k}}$ descendants as above.
However we see that, unlike the single-$t$ sector, here the way that the global $\SL(2)\times \SL(2)$ generators create states appears to
be different from that of the remaining Virasoro generators. So in the following we will study the combinatorics of the $L_{-1}$ and $\Lb_{-1}$ descendants separately from the $L_{-k},\Lb_{-\bar{k}}$ descendants for $k,\bar{k}\geq2$.  

To do this, let us rewrite the two-particle term (suppressing the overall $q^4$ for clarity) as
\be
\begin{split}
  \frac{1}{2!}Y_2 &=\frac{1+q\qb}{(1-q)(1-q^2)(1-\qb)(1-\qb^2)}=(1+q\qb)\sum_{k=0}\sum_{\bar{k}=0}p(k,2)q^k p(\bar{k},2)\qb^{\bar{k}}\\
  &=\sum_{k=0}\sum_{\bar{k}=0}p(k,2) p(\bar{k},2)q^k\qb^{\bar{k}}+\sum_{k=1}\sum_{\bar{k}=1}p(k-1,2) p(\bar{k}-1,2)q^k\qb^{\bar{k}}\;,
\end{split}
\ee
where $p(k,2)$ denotes the number of partitions of $k$ into 2 parts\footnote{Recall that, in general, $\prod_{l=0}^k\frac{1}{1-q^l}=\sum_{n=0}^\infty p(n,k)q^n$.} and where in the last line we have changed the range
of summation to more easily see the contributions to any given term of order $(k,\bar{k})$.  Noting that the value of
$p(k,2)=\left\lfloor \frac{k}{2}+1 \right\rfloor$, we can easily distinguish four different cases, depending on whether
the levels $k$ and $\bar{k}$ (or equivalently, the numbers of $L_{-1}$ and $\Lb_{-1}$ operators acting on $\ket{t}\otimes \ket{t}$)  are
even or odd. Focusing on a given term of level $q^k\qb^{\bar{k}}$ (where we are still suppressing the overall $q^4$), we find:
\be
\text{Number of states}= \left\{\begin{array}{l} 2 p(k,2) p(\bar{k},2) \quad \text{if $k$ odd and $\bar{k}$ odd}\\
                                                 2 p(k,2) p(\bar{k},2)-p(k,2) \quad \text{if $k$ odd and $\bar{k}$ even}\\
                                                 2 p(k,2) p(\bar{k},2)-p(\bar{k},2) \quad \text{if $k$ even and $\bar{k}$ odd}\\
                                                 2 p(k,2)p(\bar{k},2)-p(k,2)-p(\bar{k},2)+1 \quad \text{if $k$ even and $\bar{k}$ even}\\
\end{array}\right.
\ee
The interpretation of this counting is as follows: The number of different states we can create from $k$ $L_{-1}$ operators
and $\bar{k}$ $\Lb_{-1}$ operators is naively $2\times p(k,2)\times p(\bar{k},2)$. For instance, if $k=3$ and $\bar{k}=1$ we can
partition the three $L_{-1}$ operators in two ways (since $p(3,2)=2$): $L_{-1}^3\otimes 1$ and $L_{-1}^2 \otimes L_{-1}$. Then, we
can introduce the single $\Lb_{-1}$ operator (since $p(1,2)=1$) in two ways ($\Lb_{-1}\otimes 1$ and $1\otimes \Lb_{-1}$), to
obtain four different overall operators:
$L_{-1}^3\Lb_{-1}\otimes 1$, $L_{-1}^3\otimes \Lb_{-1}$, $L_{-1}^2 \Lb_{-1}\otimes L_{-1}$ and $L_{-1}^2 \otimes L_{-1}\Lb_{-1}$. However,
when $\bar{k}$ is even, this algorithm overcounts, as one of the partitions will be
$\Lb_{-1}^\frac{\bar{k}}2\otimes \Lb_{-1}^\frac{\bar{k}}2$ which
is symmetric. So one needs to subtract a factor of $p(k,2)$. Similarly if $\bar{k}$ is odd and $k$ is even one subtracts $p(\bar{k},2)$.
If both $k$ and $\bar{k}$ are even there are two such symmetric
partitions which need to be subtracted, but then one undercounts by the doubly symmetric state $L_{-1}^\frac{k}2\Lb_{-1}^\frac{\bar{k}}2\otimes L_{-1}^\frac{k}{2}\Lb_{-1}^\frac{\bar{k}}2$ which needs to be added.

The outcome of this reasoning is very simple: The $Y_2$ term in (\ref{ZTMGBellfinal}), which led to the third term in
the braces in (\ref{ZTMGexpandedfinal}), simply counts all the descendants of $\ket{t}\otimes \ket{t}$ by the
action of $L_{-1}$ and $\Lb_{-1}$ in all possible different ways. Bringing back the overall factor of $q^4$, a given term
at absolute level $l,\bar{l}$ in $\mathcal{B}(q,\qb)$ will
count all the different states by all independent permutations of $l-4$ $L_{-1}$ and $\bar{l}$ $\Lb_{-1}$ generators. (The states
will need to be symmetrised to account for bosonic statistics, but this will not affect the above counting based on the partitions). 
After this counting is done, it is straightforward to account for the overall prefactor, whose combinatorics are precisely
the same as in the single-particle sector. Of course, in explicitly constructing the descendants, instead of e.g.
$L_{-k}$ acting on $\ket{t}$ we need to act with the appropriate coproduct $\Delta(L_{-k})=L_{-k}\otimes 1+1\otimes L_{-k}$ on $\ket{t}\otimes \ket{t}$. To illustrate how this
works, we work out some explicit examples in section \ref{Counting}.

All the above was for the 2-particle sector, however the same holds for the higher terms in the Bell expansion as well. For instance,
the fourth term in the braces in (\ref{ZTMGexpandedfinal}) is equal to 
\be
\frac{1}{3!}Y_3 \, q^6=\frac{1+q\qb+q\qb(q+\qb)+q^2\qb^2+q^3\qb^3}{(1-q)(1-q^2)(1-q^3)(1-\qb)(1-\qb^2)(1-\qb^3)} ~q^6
\ee
where in the denominator we recognise the generating function of the partitions $p(k,3)$ and $p(\bar{k},3)$ of $k$ and $\bar{k}$
into 3 parts, while the numerator will account for the partitions with symmetric parts. Although we will not delve into a
detailed study of the various cases as for the two-particle case, we provide examples of how the counting works in the
$3$-particle case in section \ref{Counting}. We can thus confirm that the $Y_3$ term in $\mathcal{B}(q,\qb)$ correctly
counts $L_{-1}$ and $\Lb_{-1}$ descendants of $\ket{t}\otimes \ket{t}\otimes \ket{t}$. 

To conclude this section, the splitting of $Z_{TMG}$ into $\mathcal{A}(q,\qb)$ and $\mathcal{B}(q,\qb)$ and the rewriting
of $\mathcal{B}(q,\qb)$ as a Bell generating function allowed us to simply understand the counting of multi-particle states:
Each $Y_n$ term in $\mathcal{B}(q,\qb)$ contains all the $L_{-1}$ and $\Lb_{-1}$ descendants of $\ket{t}\otimes_n\ket{t}$, while
the $\mathcal{A}(q,\qb)$ prefactor contains the $L_{-k}$ and $L_{-\bar{k}}$ descendants with $k,\bar{k}\geq 2$.

\section{Bell polynomials and the Plethystic Exponential} 

In the previous section, we described how reformulating the TMG partition function as a generating function of Bell polynomials helps to elucidate the counting of multi-particle states and their descendants. We would now like to draw a parallel between these results and a mathematical tool known as the \textit{plethystic exponential}, an exponential generating function of \textit{Hilbert series}.
  
Also called \textit{Molien} or \textit{Poincar\'{e}} function, the Hilbert series is a generating function familiar in algebraic geometry for counting the dimension of graded components of the coordinate ring. Its approach has been developed and extensively used in theoretical physics, particularly in the counting of BPS states in supersymmetric gauge theory, under the so-called Plethystic Program initiated in \cite{Benvenuti:2006qr,Feng:2007ur} (earlier work in the context of duality appeared in \cite{Pouliot:1998yv}).

The plethystic exponential generates all symmetric combinations of the variables of the Hilbert series, and in the context of the Plethystic Program is used  to obtain the generating function of multi-trace operators in gauge theory from the generating function of single-trace operators. This formalism can equally well be applied to our setting of generating multi-particle states.  Following \cite{Benvenuti:2006qr,Feng:2007ur} we define the bosonic plethystic exponential $PE^{\mathcal{B}}$ as\footnote{The definition of the plethystic exponential
  in this specific form appears in \cite{Forcella:2007ps}.}  
\be
\prod^{\infty}_{n=0} \frac{1}{\left( 1- \nu q^n    \right)^{a_n}} = PE^{\mathcal{B}} [\mathcal{G}_1(q)] 
\equiv \exp \left( \sum^{\infty}_{k=1} \frac{\nu^k}{k}    \mathcal{G}_1 \left( q^k \right) \right) = 
\sum^{\infty}_{N=0} \nu^N \mathcal{G}_N (q), 
\ee 
where 
\be 
\mathcal{G}_{1} (q) = \sum^{\infty}_{n=0} a_n q^n\;. 
\ee 
Here the integer $a_n$ indicates the number of operators with dimension $n$. The plethystic exponential
takes a certain function $\mathcal{G}_1(q)$ and, through its
exponentiation, generates 
new partition functions $\mathcal{G}_N(q)$ counting all $N$-tuple symmetric products of the 
constituents of $\mathcal{G}_1(q)$, in accordance with bosonic statistics. Depending on what is being 
counted, the term $\nu$ is sometimes referred to as root coordinates, fugacities or monomials in 
weight (see e.g. \cite{Hanany:2016gbz} for a discussion).

The plethystic exponential in one variable $q$ can also be generalised to a set of variables $q_i$ \cite{Benvenuti:2006qr}:\footnote{We
  write the formula for any number of variables to facilitate comparison with \cite{Benvenuti:2006qr}, however we will immediately specialise
  to the case of two variables $q_1,q_2$, which will be sufficient for our purposes.}
\be
\begin{split}
\prod^{\infty}_{p_1,\ldots,p_m} \frac{1}{\left( 1- \nu q_1^{p_1} \cdots q_m^{p_m}   
\right)^{a_{p_1,\ldots, p_m}}} &= PE^{\mathcal{B}} \left[\mathcal{G}_1 \left(q_1,\ldots,q_m\right) \right] \\   
&\equiv \exp \left( \sum^{\infty}_{k=1} \frac{\nu^k}{k}    \mathcal{G}_1 \left( q_1^k,\ldots,q_m^k \right) 
\right) = \sum^{\infty}_{N=0} \nu^N \mathcal{G}_N (q_1,\ldots,q_m), 
\end{split}
\ee
with: 
\be
\mathcal{G}_{1} (q_1,\ldots,q_m) = \sum^{\infty}_{p_1,\ldots,p_m =0} a_{p_1,\ldots,p_m} q_1^{p_1} 
\cdots q_m^{p_m}. 
\ee
Let us now show that our Bell polynomial results are the same as what one would obtain
by following the plethystic exponential prescription in the case of two variables. We will make the
following specialisation of the variables above: 
\be \label{PEspecial}
a_{p_1,p_2} = 1 \;\;, \;\; q_1=q \;\;,\;\; q_2=\bar{q} \;\;,\;\; p_1=m \;\;,\;\; p_2=\bar{m} \;\;,\;\; 
\nu = q^2\;.
\ee
We then immediately recover the double product $\mathcal{B} \left( q, \bar{q} \right)$ in (\ref{Bpart}).
In our case, the term $\nu$ is specialised to be a monomial in weight. Then, taking: 
\be 
\mathcal{G}_{1} (q,\bar{q}) = \sum^{\infty}_{m \geq 0,\bar{m} \geq 0}    q^{m}    \bar{q}^{\bar{m}}, 
\ee 
the plethystic exponential of $\mathcal{G}_{1} (q,\bar{q})$ is 

\be PE^{\mathcal{B}} \left[ \mathcal{G}_{1} (q,\bar{q}) \right] =    \exp \left( \sum^{\infty}_{k=1} \frac{\left( 
q^2 \right)^k}{k}    \mathcal{G}_1 \left( q^k,\bar{q}^k \right) \right) \;.
\ee 
Expanding the exponential, we obtain a series in powers of $q^2$: 
\be
\begin{split}
  \mathcal{B} \left( q, \bar{q} \right) &= PE^{\mathcal{B}} \left[ \mathcal{G}_{1} (q,\bar{q}) \right] \\
  &= 1 + \mathcal{G}_1(q,\bar{q}) \left( q^2 \right) +    \frac{\mathcal{G}^2_1 (q,\bar{q}) 
  + \mathcal{G}_1 \left( q^2, \bar{q}^2 \right)}{2} \left( q^2 \right)^2 \\
&+ 
\frac{\mathcal{G}^3_1 (q,\bar{q}) + 3    \mathcal{G}_1(q,\bar{q})      \mathcal{G}_1 \left( q^2, \bar{q}^2 
  \right) + 2 \mathcal{G}_1 \left( q^3, \bar{q}^3 \right)}{6} \left( q^2 \right)^3 \\
&+ \frac{\mathcal{G}^4_1 (q,\bar{q}) + 6 \mathcal{G}^2_1 (q,\bar{q}) \mathcal{G}_1 \left( q^2, \bar{q}^2 
\right) + 3 \mathcal{G}^2_1 \left( q^2, \bar{q}^2 \right)+ 8 \mathcal{G}_1 (q,\bar{q}) \mathcal{G}_1 
\left( q^3, \bar{q}^3 \right) + 6 \mathcal{G}_1 \left( q^4, \bar{q}^4 \right)}{24}(q^2)^4    \\ &+ \cdots   
\end{split}
\ee
Finally, it is easy to see that the coefficients of $\left( q^2 \right)^N$ are equal to the Bell polynomials at 
each order, by using the identification: 
\be
(k-1)! \mathcal{G}_1 \left( q^k,\bar{q}^k \right) = g_k \left( q, \bar{q} \right). 
\ee
In analogy with the aforementioned applications, the above derivation shows that the Hilbert series $\mathcal{G}_1 {\left( q, \Bar{q} \right)}$ counts single particles, while the plethystic exponential $PE^{\mathcal{B}}\left[ \mathcal{G}_1 {\left( q, \Bar{q} \right)} \right]$ counts multi-particles, and provides further confirmation that we have reorganized $Z_{TMG}$ in a way that clearly shows the single particle and multi-particle Hilbert spaces of the logarithmic states.

We close this section by emphasising that the similarity between our method and the one of the plethystic
exponential is not unexpected. After all, Bell polynomials are a result of plethysm associated 
with multipartite exponential generating functions. However, the Bell polynomial formalism  allows us to
uncover hidden symmetry actions on the $n$-particle terms in the partition function, as we will see in Section \ref{ladder}.
Before that, however, we will demonstrate how our results match with an explicit counting of
the multi-particle states and their descendants in the dual LCFT.

\section{Explicit counting of multi-particle descendants} \label{Counting}

To illustrate how each Bell polynomial term in (\ref{ZTMGexpanded}) counts the descendants of multi-particle
states of the logarithmic partner $t$, we show how the counting works for some low-lying states. 
As indicated at the end of section \ref{Bellmultiparticle}, we will follow a two-step procedure for constructing
the multi-particle $t$ states. First, we will focus on the $\mathcal{B}(q,\qb)$ part and
identify the states corresponding to the terms in the Bell polynomial formula. As argued, these are constructed
by the symmetrised action of the $L_{-1}$ and $\Lb_{-1}$ operators on $\ket{t}\otimes_n \ket{t}$. Then, we will add the
descendants of each of these states by the action of the $L_{-k}$ and $\Lb_{-k}$ operators, with $k\geq2$,
which will correspond to the prefactor $\mathcal{A}(q,\qb)$ in (\ref{ZTMGsplit}).

\subsection{Matching of the Bell series}

Our claim is that the $n$-th term in Bell polynomial
expansion (the terms inside the square brackets in (\ref{ZTMGexpanded})) counts all the symmetrised
(due to bosonic statistics) descendants of $\ket{t}\otimes_n \ket{t}$ by
the action of $L_{-1}$ and $\Lb_{-1}$.

Recall that the (holomorphic,antiholomorphic) weight of $\ket{t}\otimes_n \ket{t}$ is $(2n,0)$. So to construct a
descendant at absolute level $(l,\bar{l})$,
we will need to partition $l-2n$ $L_{-1}$ and $\bar{l}$ $\Lb_{-1}$ generators across $n$ copies of $t$, modulo the action
of $S_n$.

\paragraph{2-particle states}

\mbox{}

We will start with the expansion of the two-particle component of the Bell formula. It gives:
\be \label{Bell2part}
\begin{split}
\frac{1}{2!}Y_2(q^2)^2=\frac{g_1^2+g_2}{2!}(q^2)^2&=q^4+(q^5+q^4\qb^1)+(2q^6+2q^5\qb^1+2q^4\qb^2)\\
&\;\;\;+(2q^7+3q^6\qb^1+3q^5\qb^2+2q^4\qb^3)+\cdots
\end{split}
\ee
The descendants of $t\otimes t$ corresponding to the expansion (\ref{Bell2part}) are\footnote{From now on we will write
  $t\otimes t$ as shorthand for $\ket{t}\otimes \ket{t}$, and similarly for higher tensor products.}
\be \label{Explicit2part}
\begin{split}
  (4,0)&: t\otimes t\\
  (5,0)&: L_{-1}t\otimes t\\
  (4,1)&: \Lb_{-1}t\otimes t\\
  (6,0)&: L_{-1}^2t\otimes t\;,\;L_{-1}t\otimes L_{-1}t\\
  (5,1)&: L_{-1}\Lb_{-1}t\otimes t, L_{-1}t\otimes \Lb_{-1}t\\
  (4,2)&: \Lb_{-1}^2t\otimes t\;,\;\Lb_{-1}t\otimes \Lb_{-1}t\\
  (7,0)&: L_{-1}^3t\otimes t\;,\;L_{-1}^2t\otimes L_{-1}t\\
  (6,1)&: L_{-1}^2\Lb_{-1} t\otimes t\;,\;L_{-1}^2t\otimes \Lb_{-1}t \;,\;L_{-1}t\otimes L_{-1}\Lb_{-1}t\\
  \cdots\;\;&
\end{split}
\ee
Here all states should be understood as completely symmetrised due to bosonic statistics, e.g.
$L_{-1}^2t\otimes \Lb_{-1}t\rightarrow\half(L_{-1}^2t\otimes \Lb_{-1}t+\Lb_{-1}t\otimes L_{-1}^2t)$. We suppress
the symmetrisation to avoid clutter.

It can be seen (also by working out examples at higher levels) that the explicit states precisely match
the expansion (\ref{Bell2part}). This is expected, of course, according to the discussion in Section \ref{Bellmultiparticle}.

\paragraph{3-particle states}

\mbox{}

To count the descendants of $t\otimes t\otimes t$ we expand
\be \label{Bell3part}
\begin{split}
  \frac{1}{3!}Y_3(q^2)^3&=\frac{g_1^3+3g_1g_2+g_3}{3!}(q^2)^3\\
  &= q^6+(q^7+q^6\qb)+(2q^8+2q^7\qb+2q^6\qb^2)+(3q^9+4q^8\qb+4q^7\qb^2+3q^6\qb^3)+\cdots
  \end{split}
\ee
According to the discussion in Section \ref{Bellmultiparticle}, these states will be descendants of $t\otimes t\otimes t$:
\be \label{Explicit3part}
\begin{split}
  (6,0)&: t\otimes t\otimes t\\
  (7,0)&: L_{-1}t \otimes t\otimes t\\
  (6,1)&: \Lb_{-1}t\otimes t\otimes t\\
  (8,0)&: L_{-1}^2t\otimes t\otimes t\;,\;L_{-1}t\otimes L_{-1}t \otimes t\\
  (7,1)&:  L_{-1}\Lb_{-1}t\otimes t\otimes t\;,\; L_{-1}t\otimes \Lb_{-1}t\otimes t\\
  (6,2)&: \Lb_{-1}^2t \otimes t\otimes t\;,\;\Lb_{-1}t\otimes \Lb_{-1}t \otimes t\\
  (9,0)&: L_{-1}^3t\otimes t\otimes t\;,\;L_{-1}^2t\otimes L_{-1}t\otimes t\;,\;L_{-1}t\otimes L_{-1}t\otimes L_{-1}t\\
  (8,1)&: L_{-1}^2\Lb_{-1}t\otimes t\otimes t\;,\;L_{-1}^2t\otimes \Lb_{-1}t\otimes t\;,\;L_{-1}\Lb_{-1}t\otimes L_{-1}t\otimes t\;,\;
  L_{-1}t\otimes L_{-1}t \otimes \Lb_{-1}t\\
  \cdots\;\;&
\end{split}
\ee
Again, these states should be symmetrised, e.g. $L_{-1}^2t\otimes \Lb_{-1}t\otimes t$ actually means:
\scriptsize
\be \nonumber
 \frac{L_{-1}^2t\otimes \Lb_{-1}t\otimes t+L_{-1}^2t\otimes t\otimes  \Lb_{-1}t+t\otimes L_{-1}^2t\otimes \Lb_{-1}t
  +\Lb_{-1}t\otimes L_{-1}^2t\otimes t+\Lb_{-1}t\otimes t\otimes L_{-1}^2t+t\otimes \Lb_{-1}t\otimes L_{-1}^2t}{6}\;.
\ee
\normalsize
As before, we find agreement with the Bell series (\ref{Bell3part}). 
Similarly, one can confirm that $Y_n$ correctly counts the higher-order descendants of the action of $L_{-1}$ and $\Lb_{-1}$
on $t\otimes_n t$.
It is important to note that the above way of producing descendants when acting on multi-particle states
cannot be written in terms of the standard Lie-algebraic coproduct $\Delta(L_{-1})=L_{-1}\otimes 1+1\otimes L_{-1}$ (and
similarly for $\Lb_{-1}$) acting on $t\otimes t$. Simply creating states with powers of the coproduct ($\Delta(L_{-1}^k)=(\Delta(L_{-1}))^k$), as we will do in the next section, would lead to only one state at each level, and would not match $\mathcal{B}(q,\qb)$. The coproduct structure behind this counting will be the subject of an upcoming publication \cite{YMStoappear}.

\subsection{Matching of the full partition function}

Having confirmed that each term in the Bell expansion correctly counts the $L_{-1}$ and $\Lb_{-1}$ descendants of $t\otimes_n t$,
we now need to ensure that the full partition function $Z_{TMG}$ can be matched to descendants of $t\otimes_nt$. However, this is immediate
to see by considering the prefactor $\mathcal{A}(q,\qb)$ in (\ref{ZTMGsplit}). Since each term in the $n$-particle
Bell expansion is multiplied by the same prefactor $\mathcal{A}(q,\qb)$, the counting will be as in the
single-particle sector, i.e. through the combinatorics of the $L_{-k}$ and $\Lb_{-\bar{k}}$ Virasoro generators, with $k,\bar{k}\geq2$.
The only difference is that to act on the product states $t\otimes_n t$ one needs to consider the appropriate combinations of
the Lie-algebraic coproduct $\Delta(L_{-k})=L_{-k}\otimes 1+1\otimes L_{-k}$.

\paragraph{2-particle states}

\mbox{}

Including the prefactor, we find the following result for the two-particle spectrum:
\be
\begin{split}
Z_{TMG}^{(tt)}&=\prod_{n=2}^\infty \frac{1}{|1-q^n|^2}\left(\frac{g_1^2+g_2}{2!}\right)q^4\\
  &=q^4+(q^5+q^4\qb)+(3q^6+2q^5\qb+3q^4\qb^2)+(4q^7+4q^6\qb+4q^5\qb^2+4q^4\qb^3)+\cdots
  \end{split}
\ee
Let us illustrate the counting by way of examples. For instance, at level $(7,0)$ two states are already included in
(\ref{Explicit2part}). The additional two states are descendants of states of lower level in (\ref{Explicit2part}):
\be
\begin{split}
  \Delta(L_{-2})\frac{L_{-1} t\otimes t+t\otimes L_{-1} t}2&=(L_{-2}\otimes 1+1\otimes L_{-2})\frac{L_{-1} t\otimes t+t\otimes L_{-1}t}2\\
  &=\frac{L_{-2}L_{-1}t\otimes t+L_{-2}t\otimes L_{-1}t+L_{-1}t\otimes L_{-2}t+t\otimes L_{-2}L_{-1}t}2
\end{split}
\ee
and
\be
  \Delta(L_{-3})t\otimes t=(L_{-3} t\otimes t+t\otimes L_{-3}t)\;,
  \ee
  where we have made the symmetrisation explicit. At level $(5,2)$, we have 3 states included in (\ref{Explicit2part}) along with:
\be
\Delta(\Lb_{-2})\frac{L_{-1}t\otimes t+t\otimes L_{-1}t}2=\frac{(\Lb_{-2}L_{-1}t\otimes t+\Lb_{-2}t\otimes L_{-1}t+L_{-1}t\otimes \Lb_{-2}t+t\otimes \Lb_{-2}L_{-1}t}2\;.
\ee
One can similarly understand the counting of higher two-particle terms as well.\footnote{Note that when considering coproducts of
  multiple generators one should take into account the compatibility of the product and coproduct, e.g. $\Delta(L_{-2}^2)=
  \left(\Delta(L_{-2})\right)^2=L_{-2}^2\otimes 1+2 L_{-2}\otimes L_{-2}+1\otimes L_{-2}^2$.}

\paragraph{3-particle states}

\mbox{}

Here the relevant expansion is
\be
\begin{split}
Z_{TMG}^{(ttt)}&=\prod_{n=2}^\infty \frac{1}{|1-q^n|^2}\left(\frac{g_1^3+3g_1g_2+g_3}{3!}\right)q^6\\
  &=q^6+(q^7+q^6\qb)+(3q^8+2q^7\qb+3q^6\qb^2)+(5q^9+5q^8\qb+5q^7\qb^2+4q^6\qb^3)+\cdots
  \end{split}
\ee  
To construct descendants of the 3-particle states, one needs to consider the appropriate composition of coproducts
of the $L_{-k}$ and $\Lb_{-\bar{k}}$ generators, with $k,\bar{k}\geq 2$, so that it acts on three copies of the Hilbert space:\footnote{Associativity of the Lie-algebraic coproduct guarantees that this is equal to the other combination we could have written,
i.e. $(1\otimes \Delta) \Delta(L_{-n})$. Similarly, there is a unique coproduct for every $n$-particle sector.}
\be
(\Delta \otimes 1) \Delta(L_{-n})=(\Delta \otimes 1) (L_{-n}\otimes 1 + 1\otimes L_{-n})=L_{-n}\otimes 1\otimes 1+1\otimes L_{-n}\otimes 1
+1\otimes 1 \otimes L_{-n}\;.
\ee

As an example, at level $(9,0)$, we have already counted 3 states in (\ref{Explicit3part}) so we require two more. They are given by
\be
\begin{split}
&(\Delta\otimes 1)\Delta(L_{-2}) \frac{L_{-1}t \otimes t\otimes t+t\otimes L_{-1}t\otimes t+t\otimes t\otimes L_{-1}t}3\quad \text{and}\\
&  (\Delta\otimes 1)\Delta(L_{-3}) t\otimes t\otimes t\;.
\end{split}
\ee
For the $(8,1)$ states, we already have four so we need to add one more:
\be
(\Delta\otimes 1) \Delta(L_{-2}) \frac{\Lb_{-1}t \otimes t\otimes t+t\otimes \Lb_{-1}t\otimes t+t\otimes t\otimes \Lb_{-1}t}3\;.
\ee
It is straightforward to proceed to higher levels and construct the explicit states that agree with the required counting. 

We conclude that the $t$ multi-particle states in $Z_{TMG}$ correctly arise from the hybrid counting we have outlined
above. To summarise, one first constructs all possible $L_{-1},\Lb_{-1}$ descendants of $t\otimes_nt$, where the action of the
generators is only constrained by overall symmetry of the multi-particle wavefunction. This agrees with the corresponding
$Y_n$ term in the Bell expansion. Then, one constructs descendants of these states in the standard way,
by acting with the appropriate coproducts of the generators $L_{-k},\Lb_{-\bar{k}}$ with $k,\bar{k}\geq 2$. This accounts for the prefactor $\mathcal{A}(q,\qb)$.

\section{A ladder action on Z$_{{TMG}}$} \label{ladder}

In this section, we return to a more detailed study of the Bell polynomial version of $Z_{TMG}$, with the goal of
uncovering some additional structure. We will see that the partition function admits a natural action of ladder
(raising/lowering) operators. From these, one can further construct an $\slrm(2)$ symmetry of the $n$-particle terms
in $Z_{TMG}$. 
 
\subsection{Monomiality principle}

The operators we are concerned with will act on  $\mathcal{B} \left( q, \bar{q} \right)$, and we will
see that they generate an $\slrm(2)$ algebra. In order to motivate the appearance of this action, 
let us briefly introduce the \textit{monomiality principle}, which is a useful tool for studying 
properties of special polynomials, such as the Bell polynomial.

The idea of monomiality is rooted in the early 1940s, when J.F. Steffensen, in a paper 
\cite{steffensen1941poweroid} that only recently received attention, suggested the concept of the 
\textit{poweroid}. A resurgence of the theory arose in the work of G. Dattoli \textit{et al}, who 
systematically made use of the principle \cite{dattoli1997evolution, dattoli1999hermite}. In 
essence, all polynomial families, in particular special polynomials, are identical, as it suffices to
transform a basic set of monomials using suitable (\textit{derivative} and \textit{multiplication}) operators
to obtain the polynomials. This result, theoretically proven in \cite{cheikh2003some} and 
\cite{cheikh2002obtaining}, is closely related to the theory of \textit{Umbral Calculus} 
\cite{roman2005umbral}, since the exponent, for instance in the monomial $x^n$, 
transforms into its ``shadow'' in the polynomial $p_n(x)$.

Let us consider the \emph{Heisenberg-Weyl} algebra, i.e. the nilpotent algebra with generators
$\hat{\mathcal{D}}$ and $\hat{\mathcal{X}}$ satisfying the commutation relations: 
\be
[\hat{\mathcal{D}},\hat{\mathcal{X}}]= 1, \hspace{0.5cm} [\hat{\mathcal{D}},1] = [\hat{\mathcal{X}},1] =0\;.       
\ee
This algebra encodes the structure of raising and lowering operators, which arise in canonical quantisation. Here
we will be interested in its uses in combinatorial physics, in relation to the monomiality principle, such as in
\cite{blasiak2006representations}.

The monomiality principle is based on the fact that a given family of polynomials $p_n(x)$ can be viewed as 
\textit{quasi-monomial} under the action of two operators $\hat{\mathcal{D}}$ and $\hat{\mathcal{X}}$, called ``derivative'' 
and ``multiplicative'' operators respectively, if it satisfies the recurrence relations: 
 
\begin{eqnarray} 
\label{eq:33} 
\hat{\mathcal{X}} p_n(x) &=& p_{n+1}(x) \nonumber \\ 
\hat{\mathcal{D}} p_n(x) &=& n p_{n-1}(x) \\ 
p_n(0) &=& 1. \nonumber   
\end{eqnarray} 
These operators can immediately be seen as raising and lowering operators acting on the polynomials 
$p_n(x)$. Eqs. (\ref{eq:33}) also imply the eigenproperty of the operator $\hat{\mathcal{X}} \hat{\mathcal{D}}$: 
\be
\hat{\mathcal{X}} \hat{\mathcal{D}} p_n(x)= n p_n(x). 
\ee
It is interesting to note that the operators $\hat{\mathcal{D}}$ and $\hat{\mathcal{X}}$ satisfy the commutation relation: 
 
\be
[\hat{\mathcal{D}},\hat{\mathcal{X}}]= \hat{\mathcal{D}} \hat{\mathcal{X}} - \hat{\mathcal{X}} \hat{\mathcal{D}} =1, 
\ee
hence displaying a Weyl algebra structure.

\subsection{Ladder action on the Bell polynomials}

Let us now focus specifically on the Bell polynomials. Motivated by a construction in \cite{riordan1968}, we 
define an operator $\hat{X}$ (the concrete version of the abstract $\hat{\mathcal{X}}$ above which is appropriate
to the Bell polynomials) as: 
\be \label{XBell}
\hat{X} = g_1 + \sum^\infty_{k=1} g_{k+1} \frac{\partial}{\partial g_k}. 
\ee
This operator acts as a multiplication operator on the Bell polynomials in $n$ variables denoted in the 
previous section as $Y(g_1,g_2,\ldots,g_n)$. For $Y_n=Y(g_1,g_2,\ldots,g_n)$, we therefore have: 
 \be
\hat{X} Y_n=    Y_{n+1}. 
\ee
It is now natural to define a second operator $\hat{D}$ as   
 \be \label{DBell}
\hat{D}= \frac{\partial}{\partial g_1}\;.
\ee
This operator acts as derivative operator on $Y_n$: 
\be
\hat{D} Y_n= n Y_{n-1}. 
\ee
Finally, the combined operator $\hat{X}\hat{D}$ acts on $Y_n$ as:  
 \be
 \hat{X}\hat{D} Y_n= n Y_n. 
\ee 
It is straightforward to verify that the operators $\hat{X}$ and $\hat{D}$ are generators of the 
Heisenberg-Weyl algebra. We will next show that these operators can be used to construct a $\slrm(2)$ algebra.

\subsection{$\slrm(2)$ action on the Bell polynomials} \label{LadderBell}

Let us consider the standard $\slrm(2)$ algebra basis $\{f,e,h\}$, satisfying
\be 
[f,e]=h, \hspace{0.5cm} [h,e]=2e, \quad [h,f]=-2f. 
\ee
Now, given a Heisenberg-Weyl algebra with generators $\hat{X}$ and $\hat{D}$ as above, it is well known
\cite{feinsilver2004representations} that one can obtain a standard $\slrm(2)$ algebra through the
definitions:
\be
f=\frac{1}{2}\hat{X}^2, \hspace{0.5cm} h= \hat{X} \hat{D} + \frac{1}{2}, 
\hspace{0.5cm} e= \frac{1}{2}\hat{D}^2 \;.
\ee

 Given the results of the previous section, one can easily verify how these $\slrm(2)$ generators  act on the Bell polynomials:   
\be
e Y_{n}= \frac{1}{2} n (n-1) Y_{n-2}\;,\quad
f Y_{n}= \frac{1}{2} Y_{n+2}\;,\quad
h Y_{n}= \left( n+ \frac{1}{2} \right) Y_n\;. 
\ee

\begin{figure}[h] 
  \begin{center}
    \begin{tikzpicture}[scale=1.6]
\draw[->,blue] (0,0.1) -- (1,0.1);
\draw[->,magenta]  (1,-0.1) -- (0,-0.1);
\draw[->,blue] (1.5,0.1) -- (2.5,0.1);
\draw[->,magenta] (2.5,-0.1) -- (1.5,-0.1);
\draw[->,blue] (3,0.1) -- (4,0.1);
\draw[->,magenta] (4,-0.1) -- (3,-0.1);
\draw[->,blue] (4.5,0.1) -- (5.5,0.1);
\draw[->,magenta] (5.5,-0.1) -- (4.5,-0.1);

\draw[->,orange] (0,0.2) ..  controls (1.6,1.6) .. (2.5, 0.2);
\draw[->,orange] (3,0.2) ..  controls (4.6,1.6) .. (5.5, 0.2);

\draw[->,cyan] (5.5,-0.2) ..  controls (4.2,-1.6) .. (3,-0.2);
\draw[->,cyan] (2.5,-0.2) ..  controls (1.2,-1.6) .. (0,-0.2);

\node at (-.25,0) {$Y_i$} edge[loop above]  ();
\node at (-.25,0.7) {$h$};
\node at (1.25,0) {$Y_{i+1}$} edge[loop above]  ();
\node at (1.25,0.7) {$h$};
\node at (2.75,0) {$Y_{i+2}$} edge[loop above]  ();
\node at (2.75,0.7) {$h$};
\node at (4.25,0) {$Y_{i+3}$} edge[loop above]  ();
\node at (4.25,0.7) {$h$};
\node at (5.75,0) {$Y_{i+4}$} edge[loop above]  ();
\node at (5.75,0.7) {$h$};
\node at (6.25,0) {$\cdots$};
\node at (-.7,0) {$\cdots$};

\node at (0.5,0.3) {$\textcolor{blue}{\hat{X}}$};
\node at (2,0.3) {$\textcolor{blue}{\hat{X}}$};
\node at (3.5,0.3) {$\textcolor{blue}{\hat{X}}$};
\node at (5,0.3) {$\textcolor{blue}{\hat{X}}$};

\node at (0.5,-0.3) {$\textcolor{magenta}{\hat{D}}$};
\node at (2,-0.3) {$\textcolor{magenta}{\hat{D}}$};
\node at (3.5,-0.3) {$\textcolor{magenta}{\hat{D}}$};
\node at (5,-0.3) {$\textcolor{magenta}{\hat{D}}$};

\node at (1.6,1.4) {$\textcolor{orange}{f}$};
\node at (4.6,1.4) {$\textcolor{orange}{f}$};

\node at (1.2,-1.4) {$\textcolor{cyan}{e}$};
\node at (4.2,-1.4) {$\textcolor{cyan}{e}$};

\end{tikzpicture}
    \caption{Ladder operators acting on the Bell polynomials $Y_i$. Note that the $\slrm(2)$ raising/lowering action
    only connects even or odd polynomials, depending on whether the initial $i$ is even or odd.} \label{FigureBell}
\end{center}     
\end{figure}
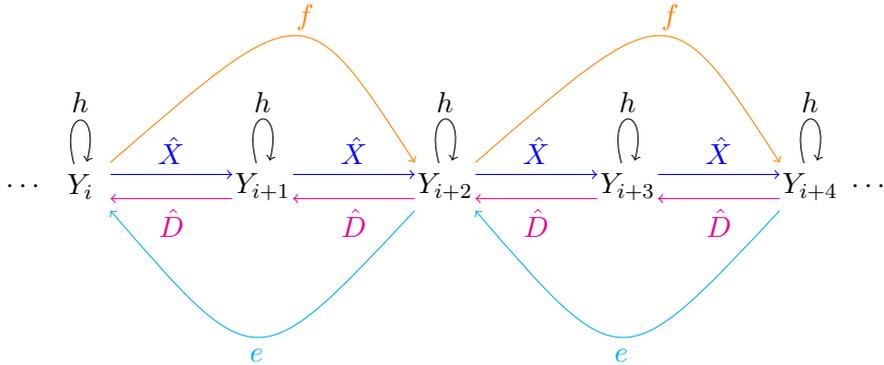

From this action we can also confirm that the algebra closes acting on the Bell polynomials:

\be
\begin{split}
[e,f] Y_n &= (ef-fe)Y_n = e \left( \frac{1}{2} Y_{n+2} \right) - f \left( \frac{1}{2}n(n-1)\right) Y_{n-2} \\   
&= \frac{1}{4} \left[ (n+2)(n+1) - n(n-1) \right] Y_n 
= \left( n + \frac{1}{2} \right) Y_n \\ 
&= h Y_n.
\end{split}
\ee 
Similarly, we can check that 
\be
[h,f] Y_n = 2 f Y_n \quad \text{and} \;\; [e,h] Y_n= 2 e Y_n.
\ee
We summarise the action of the Heisenberg-Weyl operators $\hat{X},\hat{D}$ and the $\slrm(2)$ generators $f,e,h$ in Fig. \ref{FigureBell}.

Let us emphasise that the general statements on monomiality ensure that the structures discussed in this section exist
for any polynomial family. What is specific to the Bell polynomials is the precise form of the $\hat{X}$ and $\hat{D}$ operators
in (\ref{XBell}) and (\ref{DBell}). We emphasise that this $\slrm(2)$ action relates the different multi-particle sectors
in the partition function and is not directly related to the $\SL(2)$ symmetry generated by the $L_{-1},L_0,L_{1}$ Virasoro
operators (which can only act within a given multi-particle sector).

Clearly, the raising/lowering and $\slrm(2)$ actions described above can be used to quickly produce the right counting
of states in a given multi-particle sector given the knowledge of the single-particle or another multi-particle sector. It
would be interesting to understand whether these actions can be used (for instance) to formulate consistency conditions on
putative multi-particle partition functions.

\subsection{Ladder action on the plethystic exponential} 

For completeness, we now show how from the construction of new operators satisfying the Heisenberg-Weyl 
algebra, an $\slrm(2)$ action can be obtained on successive terms of the plethystic exponential 
expansion of $PE^{\mathcal{B}} \left[ \mathcal{G}_{1} (q,\bar{q}) \right]$ (where we are of course restricting
to the very special choices in (\ref{PEspecial})).
We start by establishing some notation conventions, writing the plethystic exponential expansion as:   
 \be 
PE^{\mathcal{B}} \left[ \mathcal{G}_{1} (q,\bar{q}) \right] = PE_{(1)}    \left(    q^2 \right)^1 + 
\frac{1}{2!} PE_{(2)} \left(    q^2 \right)^2 \nonumber + \frac{1}{3!} PE_{(3)} \left(    q^2 \right)^3 + 
\ldots, 
\ee 
with:  
\be
\begin{split}
PE_{(1)} &= \mathcal{G}_1(q,\bar{q})\;,\;\; PE_{(2)} = \mathcal{G}^2_1 (q,\bar{q}) + \mathcal{G}_1 \left( q^2, \bar{q}^2 \right)\;,\\ 
PE_{(3)} &= \mathcal{G}^3_1 (q,\bar{q}) + 3    \mathcal{G}_1(q,\bar{q})      \mathcal{G}_1 \left( q^2, 
\bar{q}^2 \right) + 2 \mathcal{G}_1 \left( q^3, \bar{q}^3 \right), \cdots 
\end{split}
\ee
For the sake of clarity, we also write $\mathcal{G}_1 \left( q^k, \bar{q}^k \right) = \mathcal{G}_{1,k}$. Now, 
let us introduce an operator $\hat{\mathrm{X}}$ as: 
\begin{eqnarray} 
\hat{\mathrm{X}} = \mathcal{G}_{1,1} + \sum^\infty_{k=1} k \mathcal{G}_{1,k+1} \frac{\partial}{\partial 
\mathcal{G}_{1,k}},
\end{eqnarray} 
it is easy to see that it acts as a multiplication operator on $PE_{(k)}$ such that: 
 \begin{eqnarray} 
\hat{\mathrm{X}} PE_{(k)}=    PE_{(k+1)}. 
\end{eqnarray} 
We then define the operator $\hat{\mathrm{D}}$ as   
\begin{eqnarray} 
\hat{\mathrm{D}}= \frac{\partial}{\partial \mathcal{G}_{1,1}}, 
\end{eqnarray} 
which acts as derivative operator on $PE_{(k)}$:  
\begin{eqnarray} 
\hat{\mathrm{D}} PE_{(k)}= k PE_{(k-1)}, 
\end{eqnarray} 
and finally, the operator $\hat{\mathrm{X}}\hat{\mathrm{D}}$ that acts on $PE_{(k)}$ as: 
 \begin{eqnarray} 
\hat{\mathrm{X}}\hat{\mathrm{D}} PE_{(k)}= k PE_{(k)}. 
\end{eqnarray} 
It is straightforward to verify that the operators $\hat{\mathrm{X}}$ and $\hat{\mathrm{D}}$ are generators of the 
Heisenberg-Weyl algebra. From there, the following set of generators: 
 \begin{eqnarray} 
f=\frac{1}{2}\hat{\mathrm{X}}^2, \hspace{0.5cm} h= \hat{\mathrm{X}} \hat{\mathrm{D}} + \frac{1}{2}, 
\hspace{0.5cm} e= \frac{1}{2}\hat{\mathrm{D}}^2 
\end{eqnarray} 
will satisfy an $\slrm(2)$ algebra on $\mathcal{G}_{1,k}$.

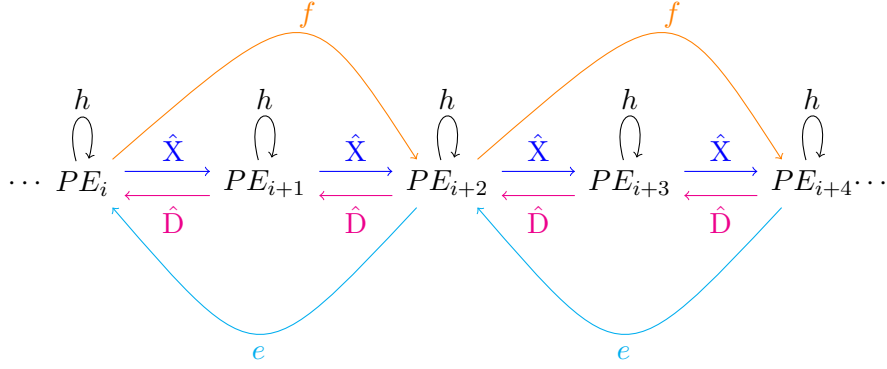
\begin{figure}[h] 
  \begin{center}
    \begin{tikzpicture}[scale=1.6]
\draw[->,blue] (0.1,0.1) -- (.8,0.1);
\draw[->,magenta]  (.8,-0.1) -- (0.1,-0.1);
\draw[->,blue] (1.7,0.1) -- (2.3,0.1);
\draw[->,magenta] (2.3,-0.1) -- (1.7,-0.1);
\draw[->,blue] (3.2,0.1) -- (3.8,0.1);
\draw[->,magenta] (3.8,-0.1) -- (3.2,-0.1);
\draw[->,blue] (4.7,0.1) -- (5.3,0.1);
\draw[->,magenta] (5.3,-0.1) -- (4.7,-0.1);

\draw[->,orange] (0,0.2) ..  controls (1.6,1.6) .. (2.5, 0.2);
\draw[->,orange] (3,0.2) ..  controls (4.6,1.6) .. (5.5, 0.2);

\draw[->,cyan] (5.5,-0.2) ..  controls (4.2,-1.6) .. (3,-0.2);
\draw[->,cyan] (2.5,-0.2) ..  controls (1.2,-1.6) .. (0,-0.2);

\node at (-.25,0) {$PE_i$} edge[loop above]  ();
\node at (-.25,0.7) {$h$};
\node at (1.25,0) {$PE_{i+1}$} edge[loop above]  ();
\node at (1.25,0.7) {$h$};
\node at (2.75,0) {$PE_{i+2}$} edge[loop above]  ();
\node at (2.75,0.7) {$h$};
\node at (4.25,0) {$PE_{i+3}$} edge[loop above]  ();
\node at (4.25,0.7) {$h$};
\node at (5.75,0) {$PE_{i+4}$} edge[loop above]  ();
\node at (5.75,0.7) {$h$};
\node at (6.25,0) {$\cdots$};
\node at (-.7,0) {$\cdots$};

\node at (0.5,0.3) {$\textcolor{blue}{\hat{\mathrm{X}}}$};
\node at (2,0.3) {$\textcolor{blue}{\hat{\mathrm{X}}}$};
\node at (3.5,0.3) {$\textcolor{blue}{\hat{\mathrm{X}}}$};
\node at (5,0.3) {$\textcolor{blue}{\hat{\mathrm{X}}}$};

\node at (0.5,-0.3) {$\textcolor{magenta}{\hat{\mathrm{D}}}$};
\node at (2,-0.3) {$\textcolor{magenta}{\hat{\mathrm{D}}}$};
\node at (3.5,-0.3) {$\textcolor{magenta}{\hat{\mathrm{D}}}$};
\node at (5,-0.3) {$\textcolor{magenta}{\hat{\mathrm{D}}}$};

\node at (1.6,1.4) {$\textcolor{orange}{f}$};
\node at (4.6,1.4) {$\textcolor{orange}{f}$};

\node at (1.2,-1.4) {$\textcolor{cyan}{e}$};
\node at (4.2,-1.4) {$\textcolor{cyan}{e}$};

\end{tikzpicture}
    \caption{Ladder operators acting between the terms of the plethystic exponential $PE_i$. The $\slrm(2)$ raising/lowering action
    only connects even or odd terms, depending on whether the initial $i$ is even or odd.} \label{FigurePE}
\end{center}     
\end{figure}

The actions of the Heisenberg-Weyl and $\slrm(2)$ operators on  $PE_{(k)}$ are illustrated in Fig. \ref{FigurePE}.

As in the Bell polynomial picture, we see that the 
single and multi-particle components of the partition function are all related via the action of
ladder (multiplication/derivative, or raising/lowering) operators.

\section{Extension to NMG and Higher-Spin TMG}

In this section, we show that the partition functions of New Massive Gravity and
Topologically Massive Spin-3 Gravity at the critical point can also be expressed in terms of Bell polynomials. 
 
\subsection{Partition function of critical New Massive Gravity} 

The partition function of New Massive Gravity at the critical point was also obtained in 
\cite{Gaberdiel:2010xv} and is given in (\ref{ZNMG}). As NMG is non-chiral (having $c_L=c_R=0$), the partition function
is symmetric in $q$ and $\qb$. 
To express (\ref{ZNMG}) in terms of Bell polynomials, we start by rewriting:
\be
\begin{split}
{{Z_{NMG}}}(q, \bar{q}) &= \prod_{n=2}^{\infty} \frac{1}{|1-q^n|^2} 
\prod_{m={\color{red}{0}}}^{\infty} \prod_{\bar{m}=0}^{\infty} \frac{1}{1- {\color{red}{q^2}} q^m 
\bar{q}^{\bar{m}}} 
\prod_{l=0}^{\infty} \prod_{\bar{l}={\color{blue}{0}}}^{\infty} \frac{1}{1- {\color{blue}{\bar{q}^2}} q^l 
\bar{q}^{\bar{l}}} \nonumber \\   
&= \mathcal{A}(q, \bar{q}) \mathcal{B}(q, \bar{q}) \bar{\mathcal{B}}(q, \bar{q}). 
\end{split}
\ee
Following essentially the same methods as for TMG, this expression can be rewritten in terms
of Bell polynomials as:
\begin{eqnarray} 
{{Z_{NMG}}}(q, \bar{q})= \prod_{n=2}^{\infty} \frac{1}{|1-q^n|^2}   
\left( \sum^{\infty}_{l=0} \frac{Y_l}{l!} \left( q^2 \right)^l \right)   \left( \sum^{\infty}_{m=0} \frac{Y_m}{m!} \left( \bar{q}^2 \right)^m    \right) .   
\end{eqnarray} 
We can further rewrite it in a binomial-type form as: 
 \be \label{ZNMGbinomial}
{{Z_{NMG}}}(q, \bar{q})= \prod_{n=2}^{\infty} \frac{1}{|1-q^n|^2} \left[ \sum_{k=0}^{\infty} 
\sum_{j=0}^k   
\frac{Y_j}{j!}   
\frac{Y_{k-j}}{(k-j)!} (q^2)^j (\bar{q}^2)^{k-j} \right]. 
\ee
To see how this is in agreement with counting of states in a non-chiral LCFT dual, let us expand the terms in the square
bracket as
\be
\begin{split}
\mathcal{B}(q,\qb)\bar{\mathcal{B}}(q,\qb) &=  1+Y_1q^2+Y_1\qb^2+\left(\half Y_2 q^4+Y_1^2 q^2\qb^2+\half Y_2 \qb^4\right)\\
&\;\;\;+\left(\frac{1}{3!}Y_3 q^6+\half Y_2Y_1 q^4\qb^2+\half Y_1Y_2 q^2\qb^4+\frac1{3!}Y_3\qb^6\right)+\cdots
\end{split}
\ee
As discussed in \cite{Gaberdiel:2010xv} and reviewed in section \ref{singlet}, the first three terms correspond to the vacuum sector, the single-particle $t$ sector and the single-particle $\tb$ sector.
These are the terms contributing to $Z^{(0)NMG}_{LCFT}$ in (\ref{Z0NMG}). The terms of the form $1/n! Y_n q^{2n}$ and their
conjugates $1/n! Y_{n} \qb^{2n}$ are descendants of $t\otimes_nt$ and $\tb\otimes_n\tb$ and will be counted exactly as for TMG.
So the first term of a new type is
\be \label{Y12}
Y_1^2 q^2\qb^2=q^2\qb^2+(2q^3\qb^2+2q^2\qb^3)+(3q^4\qb^2+4q^3\qb^3+3q^2\qb^4)+(4q^5\qb^2+6q^4\qb^3+6q^3\qb^4+4q^2\qb^5)+\cdots
\ee
Clearly, these will be descendants of the two-particle state $t\otimes\tb$. We find explicitly\footnote{We can of course symmetrise
  these states, e.g. $L_{-1} t\otimes \tb\rightarrow \half(L_{-1} t\otimes \tb+\tb\otimes L_{-1}t)$, without affecting the counting.}
\be
\begin{split}
  (2,2)&: t\otimes \tb\\
  (3,2)&: L_{-1} t\otimes \tb\;, \; t\otimes L_{-1}\tb\\
  (2,3)&: \Lb_{-1} t\otimes \tb\;,\; t\otimes \Lb_{-1} \tb\\
  (4,2)&: L_{-1}^2 t\otimes \tb\;,\; L_{-1} t\otimes L_{-1}\tb\;,\; t\otimes L_{-1}^2\tb \\
  (3,3)&: L_{-1} \Lb_{-1} t\otimes \tb\;,\; L_{-1} t\otimes \Lb_{-1} \tb\;,\; \Lb_{-1} t\otimes L_{-1} \tb\;,\; t\otimes L_{-1}\Lb_{-1}\tb\\
  (2,4)&: \Lb_{-1}^2 t\otimes \tb\;,\; \Lb_{-1} t\otimes \Lb_{-1}\tb\;,\; t\otimes \Lb_{-1}^2\tb\\
  \cdots\;\;\; &
\end{split}
\ee
We find that the number of states agrees with the counting in (\ref{Y12}). Similarly, we can look at the states arising
from
\be
\half Y_1Y_2 q^4\qb^2=q^4\qb^2+(2q^5\qb^2+2q^4\qb^3)+(4q^6\qb^2+5q^5\qb^3+4q^4\qb^4)+\cdots
\ee
which will be three-particle descendants of $t\otimes t\otimes \tb$. We can write
\be
\begin{split}
  (4,2)&: t\otimes t\otimes \tb\\
  (5,2)&: L_{-1}t\otimes t\otimes \tb\;,\;t\otimes t\otimes L_{-1}\tb\\
  (4,3)&: \Lb_{-1}t\otimes t\otimes \tb \;,\;t\otimes t\otimes \Lb_{-1}\tb\\
  (6,2)&: L_{-1}^2t\otimes t\otimes \tb\;,\; L_{-1}t\otimes L_{-1}t\otimes \tb\;,\; L_{-1}t\otimes t\otimes L_{-1}\tb\;,\;
  t\otimes t\otimes L_{-1}^2\tb\\
  (5,3)&: L_{-1}\Lb_{-1}t\otimes t\otimes \tb,L_{-1}t\otimes \Lb_{-1}t\otimes \tb,
  L_{-1}t\otimes t\otimes \Lb_{-1}\tb,\Lb_{-1}t\otimes t\otimes L_{-1}\tb,t\otimes t\otimes L_{-1}\Lb_{-1}\tb\\
  \cdots \;\;\;&
\end{split}
\ee
The counting of $4$-particle and higher sectors will proceed in a similar manner.
As for TMG, the overall prefactor in (\ref{ZNMGbinomial}) corresponds to descendants of the above states created
by the appropriate combinations of coproducts $\Delta(L_{-k})$, $k\geq 2$. In this way,
the Bell polynomial expansion has led to a better understanding of the multi-particle sector of $Z_{NMG}$.

\subsection{Partition function of critical Topologically Massive Spin-3 Gravity}

In \cite{Chen:2011vp, Chen:2011yx} and separately in \cite{Bagchi:2011vr}, critical topologically massive gravity was
generalised to
spin-3 as well as higher spins. It was seen that, analogously to the spin-2 case, the spin-3 mode becomes degenerate
with a bulk mode at the critical point $\mu l=1$, leading to the expectation that the dual CFT to higher-spin TMG
is logarithmic, and has $\Wcal$-algebra symmetry.\footnote{The two proposals differ in their choices
  of boundary conditions, and as a consequence \cite{Chen:2011vp, Chen:2011yx} find a chiral right-moving CFT, not an LCFT as in \cite{Bagchi:2011vr}.
  Furthermore, the trace part of the spin-3 field is found to be pure gauge in \cite{Chen:2011vp, Chen:2011yx} and discarded, while
in \cite{Bagchi:2011vr} it is found to be physical (and kept).} A check of this proposal was performed in \cite{Bagchi:2011td},
where the 1-loop partition function for topologically massive higher spin gravity was calculated (for arbitrary spin).
Focusing on the spin-3 case for concreteness, the result of \cite{Bagchi:2011td} takes the form:
 \be\begin{split}
{{Z^{(3)}_{TMHSG}}} (q, \bar{q})&=\prod_{n=2}^{\infty} \frac{1}{|1-q^n|^2} \prod_{m=2}^{\infty} 
\prod_{\bar{m}=0}^{\infty} \frac{1}{1-q^m \bar{q}^{\bar{m}}} \\  &\times \left[ 
\prod_{n'=3}^{\infty} \frac{1}{|1-q^{n'}|^2} \prod_{k=3}^{\infty} \prod_{\bar{k}=0}^{\infty} 
\frac{1}{1-q^k \bar{q}^{\bar{k}}} \prod_{l=4}^{\infty} \prod_{\bar{l}=3}^{\infty} \frac{1}{1-q^l 
  \bar{q}^{\bar{l}}} \right].
\end{split}
 \ee
 where we have combined the spin-2 and spin-3 parts.
Before writing the expression of ${Z^{(3)}_{TMHSG}}$ in terms of Bell polynomials, we first note that,
starting from an expression coming from gravity on the left-hand side, one obtains an 
expression featuring the $\Wcal$-algebra vacuum characters on the right-hand side. Indeed: 
\begin{eqnarray} 
{{Z^{(3)}_{TMHSG}}} (q, \bar{q})&=& \left\{ \prod_{n=2}^{\infty} \frac{1}{|1-q^n|^2} 
\prod_{n'=3}^{\infty} \frac{1}{|1-q^{n'}|^2} \right\} \prod_{m=2}^{\infty} \prod_{\bar{m}=0}^{\infty} 
\frac{1}{1-q^m \bar{q}^{\bar{m}}} \nonumber \\    &\times& \left[    \prod_{k=3}^{\infty} 
\prod_{\bar{k}=0}^{\infty} \frac{1}{1-q^k \bar{q}^{\bar{k}}} \prod_{l=4}^{\infty} 
\prod_{\bar{l}=3}^{\infty} \frac{1}{1-q^l \bar{q}^{\bar{l}}} \right] \nonumber \\ &=& \chi_0 
(\mathcal{W}_3) \times \overline{\chi_0} ({\scriptsize \mathcal{W}_3}) \prod_{m=2}^{\infty} \prod_{\bar{m}=0}^{\infty} 
\frac{1}{1-q^m \bar{q}^{\bar{m}}} \nonumber \\    &\times& \left[    \prod_{k=3}^{\infty} 
\prod_{\bar{k}=0}^{\infty} \frac{1}{1-q^k \bar{q}^{\bar{k}}} \prod_{l=4}^{\infty} \prod_{\bar{l}=3}^{\infty} \frac{1}{1-q^l \bar{q}^{\bar{l}}} \right], 
\end{eqnarray} 
where $\chi_0 (\mathcal{W}_3)$ and $\overline{\chi}_0 (\mathcal{W}_3)$ are the holomorphic and 
antiholomorphic vacuum characters of the $\Wcal_3$-algebra (\cite{Bouwknegt:1992wg}, see also \cite{Gaberdiel:2010ar}).
This property, as also pointed out in \cite{Bagchi:2011td}, fits perfectly with the expectation that the dual
theory is a $\Wcal$-LCFT.

Following the same procedure as for TMG, we can straightforwardly express the above partition function
in terms of Bell polynomials:
\be
{{Z^{(3)}_{TMHSG}}} (q, \bar{q})= \chi_0 (\mathcal{W}_3) \times \overline{\chi}_0 (\mathcal{W}_3) \times
\left[ \sum^{\infty}_{k=0} \frac{Y_k}{k!} \left( q^2 \right)^k \right]\times \left[ 
\sum^{\infty}_{l=0} \frac{Y_l}{l!} \left( q^3 \right)^l \right]\times 
\left[ \sum^{\infty}_{m=0} \frac{Y_m}{m!} \left( q^4 \bar{q}^3 \right)^m    \right]. 
\ee
The interpretation and counting of the states leading to the first and second square brackets in the above expression
will be very similar to that discussed for TMG: The $k$-th term in the first bracket will correspond to states built
upon $t\otimes_k t$, the $l$-th term in the second bracket will correspond to $l$-particle states of $w$, the logarithmic
partner of $\Wcal$, while the mixed terms will be counted similarly to the NMG case above. The counting that leads to the
third bracket is less clear to us, however, and we leave a fuller understanding
of the higher-spin case for future work.\footnote{Let us note that this term arises from the trace part of
the spin-3 field, whose presence (as mentioned above) depends on the choice of boundary conditions. }

\section{Conclusion and outlook} 

As emphasised in \cite{Grumiller:2013at}, it would be useful to exploit our knowledge of critical gravity 
theories to better understand the corresponding dual CFTs. In this paper, we made a step further towards such an  
understanding by explicitly showing how $Z_{TMG}$, derived from the gravity side in \cite{Gaberdiel:2010xv},
can be usefully recast in a compact form using Bell polynomials. These polynomials allow us to express at once
character representations of single- and multi-particles. As such, we are able to extend the checks of $Z_{TMG}$
performed in \cite{Gaberdiel:2010xv} to also include the multi-particle sector. In this way, one more of the checks
of the TMG/LCFT correspondence formulated in \cite{Grumiller:2013at} finds a positive answer.

An intriguing result is that the counting needs to be performed in a two-step process, with the combinatorics
of the $L_{-1}$ and $\Lb_{-1}$ (i.e. the global $\SL(2)\times \SL(2)$) action on $\ket{t}\otimes_n \ket{t}$ being non-standard
in that it doesn't arise from the usual Lie-algebraic coproduct $\Delta(X)=X\otimes 1+1\otimes X$. Of course,
showing that the counting works in this way does not amount to an explanation as to why. On the combinatorics side,
one could ask what type of coproduct (and associated algebraic structure) would lead to the counting of the Bell polynomial
part. This is work in progress \cite{YMStoappear}. More on the physics side, recall that the Bell polynomial expansion
arose from a rewriting of the double-product
term in (\ref{ZTMG}). In turn, the double-product term derives from the bulk massive mode, which becomes the logarithmic partner of the
boundary graviton at the critical point, so it is clearly related to the logarithmic nature of
the theory. Given that we found similar expansions for NMG and higher-spin TMG\footnote{Similar partition functions also
  arise in generalised massive \cite{Bertin:2011jk} and tricritical gravity \cite{Zojer:2012rj} so we expect the
multi-particle counting in those cases to also work out as discussed in the above.}, one could wonder whether
this type of multi-particle counting arises generically in LCFT's that could potentially be candidates for duals of
weakly-coupled 3d gravity/higher spin theories. We leave a deeper understanding of this question to future work. 

In addition, we have also shown explicitly how the  Bell polynomials are related to the plethystic exponential.
Although the existence of a relation between the two series is certainly not unexpected\footnote{A comment in this
  regard can be found, for instance, in  \cite{Lucietti:2008cv}.}, our explicit mapping could prove beneficial in future
studies in other contexts as well. An immediate benefit in our case was that the ladder action and associated $\slrm(2)$ structure
which we uncovered on the Bell polynomial side could automatically also be applied to the plethystic exponential.
It would be interesting to apply the readily available extensive technology of the plethystic exponential to further
study algebro-geometric properties of the theory, such as the moduli space of the log partners and their associated
orbits, or also the invariant nature of the log partners under group actions.

An important question is whether the TMG/LCFT correspondence can be made even more precise, in the sense of
identifying a concrete Logarithmic CFT dual to TMG. The fact that we are at $c_L=0$ might appear to
preclude a correspondence at the same level as the higher-spin/$\Wcal$-algebra CFT duality \cite{Gaberdiel:2011zw},
where $c$ can be tuned to be large. However, here the logarithmic anomaly $b_L$ takes on the role of the large-$N$
parameter, so any LCFT model which allows $b_L$ to become large could be a promising candidate. On the other hand,
even for $b_L$ small (which is the case e.g. for logarithmic minimal models \cite{Pearce:2006sz}, see also \cite{Creutzig:2013hma} for
a recent review of LCFT with a focus on models of this type)  one could perhaps still hope for
an understanding similar to that in \cite{Castro:2011zq}, where the partition functions of some, though not all, $c<1$
unitary minimal models were shown to agree with the dual gravity calculation (after summation over modular images).
A successful match was linked to the uniqueness of the modular invariant partition function of the CFT. One could similarly
hope that a careful study of partition functions for logarithmic minimal models would provide an indication of the general
features of possible bulk duals and lead to a precise correspondence.

\paragraph{Acknowledgements}

Y. M-S is grateful to D. Grumiller for discussions and communication on the partition functions of TMG and NMG, as well as T. Creutzig, M. Gaberdiel, A. Gainutdinov, A. Milas, D. Ridout, S. Wood and R. Vasseur for correspondence on Logarithmic Conformal Field Theories. KZ also acknowledges useful discussions with M. Gaberdiel and D. Grumiller. We also wish to thank the anonymous referee for constructive comments and
suggestions that have improved the quality of this work. Y. M-S was
supported through a PhD bursary by the South African National Institute for Theoretical Physics (NITheP) for
part of the duration of this work. KZ received support from the South African National Research Foundation
(grants CSUR-93735 and Incentive-103895).

\appendix

\section{Bell polynomials} \label{AppBell}

The Bell polynomials are defined through (\ref{FaadiBruno}), which we repeat here for ease of reference:
\begin{eqnarray} 
Y_{n}(g_1, g_2, \ldots, g_{n})= \sum_{\vec{k} \vdash n} \frac{n!}{k_1! \cdots k_n!}    \left( \frac{g_1}{1!} 
\right)^{k_1}    \left( \frac{g_2}{2!} \right)^{k_2} \cdots    \left( \frac{g_n}{n!} \right)^{k_n}, 
\end{eqnarray} 
where the definition of  ${\vec{k} \vdash n}$ was given in (\ref{kdashn}).
Let us see how this works up to third order in $n$. We actually start at order 2, since order 0 by convention 
gives $Y_0 = 1$, and order 1 trivially gives $Y_1 = g_1$. 
At order 2, we have two options: $\{k_1,k_2\} = \{2,0\}$ or $\{k_1,k_2\} = \{0,1\}$. Clearly, in the first 
case we have $2= 2 + 2(0)$, and in the second case, $2= 0 + 2(1)$. This gives: 
 \be
Y_2(g_1,g_2) =   \frac{2!}{2! 0!} \left( \frac{g_1}{1!} \right)^2 \left( \frac{g_2}{2!} \right)^0 + 
\frac{2!}{0! 1!} \left( \frac{g_1}{1!}\right)^0 \left( \frac{g_2}{2!}\right)^1 \nonumber = g_1^2 + g_2. 
\ee 
At order 3, we have three options: $\{k_1,k_2, k_3\} = \{3,0,0 \}$, $\{k_1,k_2, k_3\} = \{1,1,0 \}$ and 
$\{k_1,k_2, k_3\} = \{0,0,1 \}$. This gives: 
 \be
\begin{split}
Y_3(g_1,g_2, g_3) &= \frac{3!}{3! 0! 0!} \left( \frac{g_1}{1!} \right)^3 \left( \frac{g_2}{2!} \right)^0 \left( 
\frac{g_3}{3!} \right)^0 + \frac{3!}{1! 1! 0!} \left( \frac{g_1}{1!} \right)^1 \left( \frac{g_2}{2!} \right)^1 
\left( \frac{g_3}{3!} \right)^0 \\
&\;\;+ \frac{3!}{0! 0! 1!} \left( \frac{g_1}{1!} \right)^0 \left( \frac{g_2}{2!} \right)^0 \left( \frac{g_3}{3!}  \right)^1 \\   
&= g_1^3 + 3 g_1g_2 + g_3 . 
\end{split}
\ee  
Similarly, one finds 
\be
Y_4(g_1,g_2,g_3,g_4)=g_1^4+6g_1^2g_2+3g_2^2+4g_1 g_3+g_4\;,
\ee
\be
Y_5(g_1,g_2,g_3,g_4,g_5)=g_1^5+10g_1^3g_2+15g_1g_2^2+10 g_1^2 g_3+10 g_2 g_3+5g_1 g_4+g_5\;.
\ee
and so on (higher Bell polynomials can easily be generated using computer algebra).

Having these explicit expressions it is easy to check the action of the raising and lowering operators discussed in section \ref{LadderBell}.
For instance:
\be
\begin{split}
\hat{X} Y_3&=\left(g_1\!+\!g_2\frac{\p}{\p g_1}\!+\!g_3\frac{\p}{\p g_2}\!+g_4\!\frac{\p}{\p g_3}\right)(g_1^3+3g_1 g_2+g_3)=
g_1^4\!+\!6g_1^2g_2\!+\!4g_1g_3\!+\!3g_2^2\!+\!g_4=Y_4,\\
\hat{D} Y_3&=\frac{\p}{\p g_1} (g_1^3+3g_1 g_2+g_3)=3g_1^2+3 g_2=3 Y_2\;.
\end{split}
\ee

Each coefficient in a given Bell polynomial $Y_n$ corresponds to the number of partitions of a set of $n$ distinguishable
elements into subsets, whose number is given by the number of factors in each term. The
length of each subset is given by the index of $g_i$. Looking at $Y_4$, for example, one sees that
there is:
\begin{itemize}
\item One partition of $\{ABCD\}$ into a single set of length 4 (the original set),

\item 4 partitions
    into two subsets of lengths 1 and 3: [\{A\},\{BCD\}], [\{B\},\{ACD\}], [\{C\},\{ABD\}], [\{D\},\{ABC\}],

  \item 3 partitions into two subsets of length 2 and 2: [\{AB\},\{CD\}], [\{AC\},\{BD\}], [\{AD\},\{BC\}],

  \item 6 partitions into three subsets of length 1,1, and 2: [\{A\},\{B\},\{CD\}], [\{A\},\{C\},\{BD\}],

    [\{A\},\{D\},\{BC\}], [\{B\},\{C\},\{AD\}], [\{B\},\{D\},\{AC\}], [\{C\},\{D\},\{AB\}],

\item and finally one partition into four subsets of length 1,1,1,1: [\{A\},\{B\},\{C\},\{D\}]. 
\end{itemize}
So Bell polynomials naturally appear whenever one is counting partitions, and it is not surprising that they arise
in situations like ours where to construct descendants of multi-particle states one has to essentially partition
a certain number of raising operators to act on the state $\ket{t}\otimes_n\ket{t}$ (which would be a highest-weight
state in non-logarithmic theory) in all possible ways. 
 
For more information on the Bell polynomials, the reader is referred to one of the many books
concerned with the theory of partitions, such as \cite{riordan1968,theoryofpartitions}.

\bibliography{LCFT} 
\bibliographystyle{utphys} 
 
\end{document}